\documentclass[numberedappendix, apj, iop]{emulateapj}

\usepackage{graphicx,amsmath,amsfonts,amssymb}
\usepackage[usenames,dvipsnames]{color}
\usepackage[backref=none]{hyperref} 

\definecolor{MyDarkBlue}{rgb}{0,0.1,0.7}
\hypersetup{pdfborder={0 0 0},colorlinks,breaklinks=true,
  urlcolor={blue},citecolor={MyDarkBlue},linkcolor={magenta}}

\begin{document}

\title{The Vela Pulsar With an Active Fallback Disk} 
\slugcomment{Submitted to ApJ}

\author{G\"{o}k\c{c}e \"{O}zs\"{u}kan\altaffilmark{1}, K.Yavuz Ek\c{s}i\altaffilmark{1,2,*}, Valeri Hambaryan\altaffilmark{2}, \\ 
Ralph Neuh\"{a}user\altaffilmark{2},  Markus M.\ Hohle\altaffilmark{2,**}, Christian Ginski\altaffilmark{2,***}, 
Klaus Werner\altaffilmark{3}}

\altaffiltext{1}{\.{I}stanbul Technical University, Faculty of Science and Letters, Department of Physics, Maslak 34469, \.{I}stanbul Turkey} 
\altaffiltext{2}{Astrophysikalisches Institut und Universit{\" a}ts-Sternwarte, Universit{\" a}t Jena, Schillerg{\" a}{\ss}chen 2-3, 07745 Jena, Germany}
\altaffiltext{3}{Institute for Astronomy and Astrophysics, Kepler Center for Astro and Particle Physics, Eberhard Karls University, Sand 1, 72076, T\"{u}bingen, Germany}

\altaffiltext{*}{\href{mailto:eksi@itu.edu.tr}{eksi@itu.edu.tr}}
\altaffiltext{**}{Present address: Gene Center of the LMU Feodor-Lynen-Strasse 25 81377 Munich Germany}
\altaffiltext{***}{Present address: P.O. Box 9513  NL-2300 RA  Leiden, The Netherlands}


\begin{abstract}
Fallback disks are expected to form around young neutron stars. 
The presence of these disks can be revealed by their blackbody spectrum in the infrared, optical and UV bands. 
We present a re-reduction of the archival optical and infrared data of the Vela pulsar, together with the existing infrared and UV spectrum of Vela, and model  their unpulsed components with the blackbody spectrum of a supernova debris disk. We invoke the  quiescent disk solution of Sunyaev and Shakura for the description of the disk in the propeller stage and find the inner radius of the disk to be inside the light cylinder radius. We perform a high resolution X-ray analysis with XMM-Newton and find a narrow absorption feature at 0.57 keV which can be interpreted as ${\rm K}_{\alpha}$ line of He-like oxygen (OVII). The strength of the line indicates an element over-abundance in our line of sight exceeding the amounts that would be expected from ISM. The  spectral feature may  originate from the pulsar wind nebula and as being partly caused by the reprocessed X-ray radiation by the fallback disk. We discuss the lower-than-3 braking index of Vela as partially due to the contribution of  the propeller torques. Our results suggest that the pulsar mechanism can work simultaneously with the propeller processes and that the debris disks can survive the radiation pressure for at least $\sim 10^4$ years. As Vela is a relatively close object, and a prototypical pulsar, the presence of a disk, if confirmed, may indicate the ubiquity of debris disks around young neutron stars.
\end{abstract}



\section{Introduction}

A neutron star is born from the collapse of the core of a high mass star. The outer layers of the star are ejected as a supernova explosion, a fraction of the ejected mass may fall back \citep{col71,zel72} and, given the progenitor star had angular momentum, may form a disk \citep{mic81,mic88,per+14}. The mass and angular momentum of this disk, as well as the magnetic moment and initial spin of the neutron star are the initial parameters of the evolving neutron star-fallback disk system \citep*{cha+00,cha00,alp01,eks03,eks+05,ert+09}. According to the fallback disk model, the diverse families of young neutron stars are astrophysical manifestations resulting from the interplay of the magnetosphere of the star with the surrounding disk at different epochs \citep{alp01}.

The first debris disks around neutron stars were detected \citep{wan+06,kap+09} around anomalous X-ray pulsars which commonly are assumed to be magnetars \citep{dun92,tho96}. 
\citet{wan+06} presumed a passive irradiated disk well outside the light cylinder radius $R_{\rm L}=c/\Omega_{\ast}$ where $\Omega_{\ast}$ is the angular velocity of the neutron star. A larger portion of the spectrum was later explained by \citet{ert+07} assuming an active disk accreting onto the neutron star. Recent detection of excess emission  and an absorption line in RX J0720.4$-$3125 might be due to the existence of circumstellar material  around  this radio quiet, thermally emitting isolated neutron star \citep{ham+09}.

 A fallback disk model for rotationally powered objects was first suggested by \citet{mic81} to address a number of issues such as the braking index \citep[see also][]{men+01,cal+13,liu+14}, pulsar nulling and drifting subpulses \citep*[see also][]{cor08,mot+13a,mot+13b}. Since then further theoretical motivations for the presence of disks around young neutron stars had been presented. \citet{mic88} suggested that a fallback disk shrouds the putative pulsar in SN 1987A, in order to address the lack of pulsations.  \citet{mey92} suggested that the observed lightcurve for SN 1987A might include contributions due to accretion from 
a disk around the remnant \citep[see also][]{min93}. \citet*{lin+91} studied formation of planets within fallback disks. \citet*{per+00}  computed the emission expected from fallback disks showing that the excess emission in 
the R and I bands observed for the pulsar PSR 0656+14 is broadly consistent with emission from a disk. \citet*{mar+01age} showed that the discrepancy between the supernova age and the characteristic age, $\tau_{\rm c} \equiv P/2\dot{P}$,  of pulsar B1757-24 can be explained
by the presence of a debris disk acting a spin-down torque in addition to the magnetic dipole radiation torque. \citet*{alp+01} investigated
the evolution,  on the $P-\dot{P}$ diagram,  of pulsars acted by the spin-down torque from a fallback disk assisting the magnetic dipole radiation torque.   \citet{qia+03} suggested that the long-term, highly periodic and correlated variations in both the pulse shape and the
slow-down rate of two pulsars, PSR B1828-11 and PSR B1642-03, are due to their interaction 
with a surrounding fallback disk. \citet{bla04} suggested that the jets associated with the Crab and the Vela pulsars  might indicate the presence of fallback disks in these systems.

Vela (PSR B0833$-$45) is the first rotationally powered pulsar discovered \citep{lar+68} in the southern
hemisphere. Its spin period ($P=2\pi/\Omega_{\ast}=89$ msec) and period derivative ($\dot{P} = 1.25 \times 10^{-13}$ s/s) \citep{rad+69} indicate a characteristic age of $\tau_{\rm c} \equiv P/2\dot{P}=11.4$~kyr, placing Vela between the young Crab-like 
and the middle-aged pulsars. 
The spin-down luminosity of Vela is
$\dot{E}=4\pi^2 I\dot{P}/P^3=6.9\times 10^{36}I_{45}$ ergs s$^{-1}$ where $I_{45}$ is the moment of inertia of the star in units of $10^{45}$ g cm$^2$  and, assuming pure magnetic dipole radiation, its  magnetic field strength is $B_{\rm MDR} \cong 3 \times 10^{12}$ G.

 The optical counterpart of Vela \citep{las76} shows pulsations \citep{wal+77}. The high pulse fraction of the optical pulsations \citep{mig01} as well as the relatively high polarization \citep{wag00,mig+07pol} suggest a magnetospheric origin for the object's output in this band. Yet \citet{pet+78} studied the pulsed and unpulsed components of the Vela pulsar and found the unpulsed background amounts to a few percent (3.6 percent in the Crab case) of the peak intensity and 50 percent in the mean pulsed light implying an excess emission that may not originate from the magnetosphere.  \citet{nas+97} showed that the UBVR fluxes of the Vela pulsar are well above the Rayleigh-Jeans extrapolation of the ROSAT soft X-ray spectral fits by \citet{oge+93}, suggesting the origin of this excess emission is not the thermal radiation from the neutron star surface.

Recently, \citet{dan+11} presented the dereddened infrared, optical and UV spectrum of Vela by compiling data from near-infrared  \citep{shi+03}, optical \citep{mig+07opt} and UV \citep{rom+05} observations.  As possible origins of the  excess emission in the IR, they discussed the possibilities such as
unresolved pulsar nebula structures, pulsar magnetospheres and X-ray irradiated fallback disks. 
For the latter they considered a passive irradiated disk with an inner radius of 
$R_{\rm in} \simeq 1.95 \times 10^{10} \; {\rm cm}$
which is beyond the light cylinder radius $R_{\rm L}=4.26 \times 10^8\, {\rm cm}$ for Vela, well in the radiation zone.
It is hard to justify the stability of the magnetopause of a disk in the radiation zone given that the magnetic pressure scales as $R^{-2}$ whereas the ram pressure in the disk scales as $R^{-5/2}$ meaning that ram pressure would be larger inside the equilibrium point whereas the opposite is required for stability \citep[see e.g.][]{lip92,eks05}. More recently, \citet{zyu+13} presented an update of the data  of \citet{dan+11} and associated the excess emission---with respect to the extrapolation of the UV continuum---of Vela in the J$_s$HK$_s$ bands with its counter-jet.

Here, we model the  unpulsed spectrum of Vela in the infrared, optical and UV bands by the presence of a  fallback disk well inside the light cylinder radius. We provide further evidence for the presence of the disk as indicated by the spectral features resulting from the reprocessing of the X-ray radiation of the central source by the fallback disk. In \S \ref{sec-photometry} we present our reanalysis of the photometry. In \S \ref{sec-model} we introduce our model for reproducing the spectrum of the disk around Vela. In \S \ref{sec-spectrum} we present our spectral fit to the spectrum. In \S \ref{sec-X-ray} we describe our X-ray analysis and present another evidence for the presence of circumstellar material. We discuss the implications of our results in \S \ref{sec-discuss} and present our conclusion in \S \ref{sec-conclude}. We provide an appendix describing the evolution of the disk in terms of  its initial mass and angular momentum.

\section{Photometry of the Vela pulsar}
\label{sec-photometry}

\subsection{Observations and data reduction}

All the data reduced by us are taken from
public data archives. 

\paragraph{Very Large Telescope data}
The Vela pulsar was observed with the Infrared Spectrometer And Array Camera \citep[ISAAC;][]{moo+98} on the VLT on three different occasions on 2000-12-14, 2000-12-15 and 2001-01-05. Of these observations the first one was performed in the J$_{s}$ filter, while the latter two were performed with the H filter. All data were taken in the jitter imaging mode to enable an accurate subtraction of the bright infrared background. The ESO-Eclipse jitter routine was used for reduction of the data. All individual exposures were first flatfielded with twilight flatfield images taken in the same night. Consecutive images were then subtracted to remove the sky background and dark current. The final difference images were then shifted and median combined. To ensure removal of bad pixels and cosmic rays the highest and lowest three pixel values were always rejected in the combination process.

Upon inspection of the final reduced images, we found that the Vela Pulsar was detected in all of them as was already reported in \citet{shi+03}. However, the H band image of December 2000 had an overall shorter exposure time and hence the signal-to-noise of Vela was lower than in the image of 2001. We thus decided to discard the earlier observation and analyse only the later one in detail. Details of the observations are listed in Table~\ref{tab: obs}.

\paragraph{Hubble Space Telescope data}
The Vela pulsar was observed with the HST Wide-Field Planetary Camera 2 \citep[WFPC2;][]{tra+94} in the optical F814W, F675W and F555W broad band filters, which roughly correspond to Johnson I, R and V-band. The Vela Pulsar was always located in the higher resolution Planetary Camera. Data reduction included flat-fielding, dark current subtraction and geometric distortion correction as well as cosmic ray removal, which were all performed by the HST archive pipeline.

\paragraph{Gemini-South data}

Observations of the Vela Pulsar were undertaken at the Gemini-South telescope with the new Gemini South Adaptive Optics Imager \citep[GSAOI;][]{mcg+04} on 2013-01-30. A total of 20 dithered exposures were taken with an exposure time of 100\,s each. Upon visual inspection of the raw data, we noticed that the AO loop was not closed in one case and hence we discarded the respective exposure, leaving us with 19 exposures total. We again utilized ESO-Eclipse for reduction of the data. First, the offsets of all images were determined as input for the reduction routine, then all images were flat-fielded and sky-subtracted, as described for the VLT/ISAAC data. In the final reduced image the Vela pulsar is very well detected.

\subsection{Photometric calibration and measurements}

\paragraph{Very Large Telescope data}

The zero point of the ISAAC images were determined by imaging standard stars. The subsequent analysis was done by \citet{shi+03}, who determined a zero point of 24.81\,$\pm$\,0.04\,mag in J$_{s}$-band and 24.56\,$\pm$\,0.07\,mag in H-band. We adopted these values for our analysis. Since especially in the H-band image there is some structure close to the pulsar visible which is attributed to the counter-jet \citep{shi+03}, we decided to use PSF fitting photometry rather than aperture photometry in order not to overestimate the flux of the pulsar. For the analysis we utilized IDL and Starfinder.

We first constructed an empirical PSF utilizing multiple stars visible in the field of view along with the pulsar. We then normalized this PSF and fit it to all sources in the image in order to obtain accurate positions. In order to take the variable background into account we used Starfinder to create a background image by locally fitting a slanting plane.
The background image and the extracted PSF were then combined and fit to each previously detected source. 
The fitting process takes the estimated background noise into account in computing the uncertainties for 
the fitting results. As a result, we obtained the flux for each detected source in the image. We then converted flux into the magnitude in the respective band by
\begin{equation}
	m = -2.5 \, \log_{10}({\rm flux}/t_{\rm exp}) + m_{\rm Zero}
\end{equation}   
where $t_{\rm exp}$ is the exposure time of the individual exposures. In addition, we took atmospheric extinction and interstellar reddening into account. To correct for the atmospheric extinction, we utilized the standard extinction over the Cerro Paranal in H and J$_{s}$-band which in both cases is 0.06\,mag/airmass \footnote{\url{http://www.eso.org/sci/facilities/paranal/instruments/isaac/tools/imaging_standards.html}}. We then used the average airmass as reported in the image header, which was 1.068 for the J$_{s}$-band observation and 1.072 for the H-band observation.

For interstellar reddening we adopted a value of A$_{V}$\,=\,0.18\,mag \citep{shi+03}. This was converted into an extinction in the respective bands by the transformations provided in \citet{rie+85}. We calculated an interstellar extinction in J$_{s}$-band of 0.051\,mag and in H-band of 0.032\,mag. The final de-reddened and atmospheric extinction corrected magnitudes of the Vela pulsar can be found in Table~\ref{tab: phot}. The given uncertainties include the uncertainties of the zero point, the uncertainty of the PSF photometry, as well as an assumed 10\,\% uncertainty of the extinction and reddening corrections.

To check the reliability of our computed magnitudes, we compared for several stars within the same field of view as the pulsar, the 2MASS \citep{skru06} magnitudes with our own results. 
Specifically, we compared our results with 2MASS magnitudes for 11 random stars in both filters. In the J$_s$-band the mean absolute deviation between the two measurements was 0.069 mag, while the mean uncertainty of the utilized 2MASS magnitudes was 0.091 mag. Thus the deviations in case of the Js-band were not significant. In the H-band the mean absolute deviation was larger and amounted to 0.153 mag. This is larger than the mean uncertainty of the utilized 2MASS magnitudes, which was 0.116 mag in this case. However, the combined mean uncertainty of the 2MASS magnitudes and our own measurements was 0.196 mag, i.e. the 1 $\sigma$ uncertainty areas are overlapping. We thus decided that this mean absolute deviation was acceptable. 
In addition, we used the extracted Starfinder PSF and background image to subtract the Vela pulsar from the image. We then inspected the images to ensure that we do not overestimate the peak flux of the pulsar PSF and that we do not fit the flanks of the PSF to artefacts stemming from the counter-jet feature. In both cases (J$_{s}$ and H-band) the subtraction was clean and the counter-jet feature remained after the pulsar PSF was subtracted.

Our computed magnitudes in the J$_{s}$ and H-band are both slightly brighter than reported in \citet{shi+03} ($\Delta$J$_{s}$\,=\,0.40\,mag and $\Delta$H\,=\,0.28\,mag). We do not know what causes this difference, although it might be that they slightly overestimated the background flux which they subtracted. We are, however, confident in the results presented here.

\paragraph{Hubble Space Telescope data}

There were no standard stars imaged along with the HST observations of the Vela pulsar. Thus, the zero point of the Planetary Camera was determined from header data. Specifically we used
\begin{equation}
m_{\rm zero} = -2.5 \log_{10}({PHOTFLAM}) + {PHOTZPT}
\label{hst-zeropoint}
\end{equation}
wherein $PHOTFLAM$ and $PHOTZPT$ are header keywords. $PHOTFLAM$ represents the flux of a source with constant flux per unit wavelength which yields a count rate of 1\,$DN/s$. The zero points which are determined in this way are typically accurate down to 0.02\,mag for the F555W and F814W filters and down to 0.04\,mag for the F675W filter \citep{hey04}. Our results are listed in Table~\ref{tab: obs}. It should be stressed that the zero points 
calculated in this way are in the STmag system, which is based on a constant flux density per unit wavelength rather than the flux density of Vega.

Since the HST PSFs are generally undersampled, PSF fitting photometry is challenging. We tried different approaches consisting of building reference PSFs with the TinyTim software package and also by extracting reference PSFs directly from the imaging data with Starfinder. Fitting results in both cases varied by up to $\sim$0.3\,mag. The variations were even more prominent over different observation epochs taken with the F555W filter where magnitude differences of up to $\sim$0.6\,mag were calculated. This was not easily resolvable, without resampling of the original images. We do not, however, feel entirely confident with resampling the images, since this might affect the recorded fluxes in a not easily predictable way. We thus decided to use aperture photometry for the HST images.

All aperture photometry measurements were executed using the Aperture Photometry Tool \citep{lah+12}. Aperture radii were chosen such that only the pulsar was in the aperture and in all cases were set to either 7 or 8 pixels. Care was taken to exclude all bright sources or structures from the background estimation region, in order not to overestimate the background flux. Bright sources in the images were used to calculate aperture corrections. Correction factors were between 1.02 and 1.07. The final results are shown in Table~\ref{tab: phot}.

All apparent magnitudes were corrected for interstellar extinction assuming an A$_{V}$ of 0.18\,mag and using the transformations specific for HST filter bands provided in \citet{holtz95}. The correction values are A$_{F555W}$\,=\,0.158\,mag, A$_{F675W}$\,=\,0.123\,mag and A$_{F814W}$\,=\,0.095\,mag.

Our results differ slightly from the results given by \citet{mig01}. 
In general our magnitudes are $\sim 0.2 - 0.3$\,mag brighter. 
Since \citet{mig01} did not describe in detail how they performed their aperture photometry, we can not determine the origin of these slight differences.

\paragraph{Gemini-South data}
The photometry of the Gemini-South data was in general performed analogue to the photometry of the VLT/ISAAC data. The zero point of the detector was, however, calculated from the images of the standard star Nr. 9132 from the catalogue by \citet{per+98}. This standard star was imaged in the same night as the science data as part of the GSAOI baseline calibration plan. It was located on the same GSAOI detector quadrant as the Vela pulsar, to exclude systematic offsets between the different quadrants. The standard star was imaged in one exposure (20\,s) with three additional exposures slightly offset, which we used to construct a sky image for background subtraction. In addition, we used the IRAF task \textit{fixpix} in the \textit{NOAO} package to construct a bad pixel mask from the flat-fielded sky image. Consequently the standard star image was flat-fielded, background subtracted and bad pixels were removed by interpolation over neighbouring pixels.

Since the standard star was taken without the AO loop closed (i.e.\ under natural seeing conditions) and is not in a crowded field, we decided to use aperture photometry to determine the zero point. For this purpose we utilized again the Aperture Photometry Tool. The size of the aperture was chosen based on a curve of growth diagram determining the size at which the flux in the aperture is no longer increasing after background subtraction. Given the magnitude of the standard star, we calculated a zero point of 25.3318\,$\pm$0.0063\,mag, taking an average atmospheric extinction at Cerro Pach\'{o}n of 0.065\,mag/airmass in K$_{s}$-band into account. This is slightly larger than the average zero point of 25.17\,mag given on the GSAOI project webpage\footnote{\url{http://www.gemini.edu/?q=node/11920}}. 
However, this is the average over all four detectors of the GSAOI instrument. Thus it is not contradicting our result.

The resulting magnitude of the Vela pulsar in the K$_{s}$-band was then calculated using again IDL and Starfinder. 
We considered an interstellar extinction of 0.02\,mag in the K$_s$-band \citep{rie+85}. In addition, we corrected for an atmospheric extinction at an average airmass of 1.037. The final de-reddened and atmospheric extinction corrected magnitude can be found in Table~\ref{tab: phot}. This magnitude is in very good agreement ($\Delta$m$_K$ = 0.008 mag) with the value derived by \citep{zyu+13} from the same data set for the Vela pulsar.

\subsection{Conversion from magnitudes to fluxes}

\paragraph{Ground-based data}
All ground-based photometry was performed in the standard Vega-based magnitude system. Thus in order to calculate the total integrated flux of the Vela pulsar in the respective filter bands we had to calculate the integrated flux of Vega in the same filter bands. 
For this purpose we used a flux calibrated spectrum of Vega\footnote{\url{ftp://ftp.stsci.edu/cdbs/current_calspec/ascii_files/alpha_lyr_stis_005.ascii}} taken with the HST Space Telescope Imaging Spectrograph \citep[STIS;][]{boh04}. In addition, filter transmission curves for the utilized bands were obtained from the ISAAC\footnote{\url{http://www.eso.org/sci/facilities/paranal/instruments/isaac/inst/isaac_img.html}} 
and GSAOI\footnote{\url{http://www.gemini.edu/?q=/node/11513}} instrument web pages. 

Typically, the filter transmission curves had a lower resolution than the Vega spectrum hence we used piecewise linear interpolation to match both resolutions. The filter transmission curve was then convolved with the relevant part of the Vega spectrum to compute the spectral flux density distribution of Vega. To compute the flux of Vega, we then integrated over all frequencies utilizing Simpson's law for numerical integration. Given the flux of Vega, we calculated the integrated flux of the Vela pulsar with
\begin{equation}
F_{\rm int} = F_{\rm Vega} \cdot 10^{-0.4 m}
\label{flux}
\end{equation}
where $m$ is the magnitude of the Vela pulsar in the respective filter. In order to compute the flux at a specific frequency, we then proceeded to calculate the effective frequency with
\begin{equation}
	\nu_{\rm eff} = \frac{\int\nu T(\nu) V\!g(\nu) d\nu}{\int T(\nu) V\!g(\nu) d\nu}
	\label{nueff}
\end{equation}
and the effective width $w_{\rm eff}$ of the filter the observation was taken with
\begin{equation}
	w_{\rm eff} = \frac{\int T(\nu) d\nu}{\max(T(\nu))}
	\label{weff}
\end{equation}
where $T(\nu)$ is the transmission curve of the filter and $V\!g(\nu)$ is the Vega spectrum. Given these quantities, we calculated the flux at a specific frequency with
\begin{equation}
	\nu F_{\nu} = \frac{F_{\rm int}}{w_{\rm eff}} \nu_{\rm eff}.
	\label{nufnu}
\end{equation}
The final results are shown in Table~\ref{tab: phot}. The uncertainties of $\nu F_{\nu}$ include the uncertainties of the determined apparent magnitudes, as well as the uncertainty of the effective frequency given by the effective width of the respective filter.

\paragraph{HST data}
Conversion for HST data was in principle executed the same way as for ground-based data, with the exception that no Vega spectrum was utilized. HST magnitudes and zero points are calculated in the STmag system which is not Vega-based, but is based on constant spectral flux density per unit wavelength. Specifically, the flux density is $F_{\lambda} = 3.63 \times 10^{-9}\, {\rm erg}\, {\rm s}^{-1}\, {\rm cm}^{-2}\, \AA^{-1}$, which corresponds to the flux density of Vega at 5492.9\,$\AA$. This constant spectrum was converted to frequencies and was then utilized in the same way the Vega spectrum was utilized for the ground-based data. Results are listed as well in Table~\ref{tab: phot}.

\section{Modelling The Blackbody Spectrum of the Fallback disk}
\label{sec-model}

Predictions for the spectral energy distribution (SED) of disks around young neutron stars were given by \citet{per+00} and more recently by \citet{wer+07a}, \citet{wer+07b} and \citet{yan+12}. The model we present 
here emphasizes the importance of the propeller \citep{ill75,dav79,dav81,lov+99} boundary condition  on the spectrum. 

The realization of the propeller stage in fallback disks may be rather different than its onset in binary systems.
Fallback disks are not tidally torqued as disks in binary systems and can expand freely to take up the extra angular momentum added to the disk by the magnetosphere of the neutron star in the propeller stage.
In binary systems the mass flux within the disk is determined by the amount of material supplied by the companion. Fallback disks are not continuously fed and the mass flux within the disk is determined by the conditions at the inner boundary, i.e.\ whether the centrifugal barrier is at work or not. A self-similar solution for the disk evolution with no mass flux ($\dot{M}_{\rm d}=0$) but  freely expanding outer boundary was presented by \citet{pri74} \citep[see also][]{pri91}. A steady-state version of this solution is the ``dead disk'' solution of \citet{sun77} (hereafter SS77). To avoid confusion with dead zones thought to exist in protostellar disks,  \citet{dan10} called disks described by the SS77 solution as `quiescent disks'.

The specific intensity at frequency $\nu $ of a blackbody at temperature $T$
is
\begin{equation}
I_{\nu }=\frac{2h\nu^3/c^2}{{\rm e}^{h\nu /k_{\rm B}T}-1}
\label{bb}
\end{equation}
where $h$ is Planck's constant, $k_{\rm B}$ is Boltzmann's constant and 
$c$ is the speed of light.
The flux at frequency $\nu $ from the disk is
\begin{equation}
F_{\nu }=\frac{\cos i}{d^2}\int_{R_{\rm in}}^{R_{\rm out}}I_{\nu} 2\pi RdR  
\label{F_nu}
\end{equation}
where $d$ is the distance and $i$ is the inclination angle of the normal of
the disk with the observer.

The energy radiated by the disk has two contributions: the viscous dissipation
due to differential rotation of the shearing layers with 
enhanced  turbulent viscosity and reprocessed emission due to irradiation of the X-ray luminosity of the central object.

\subsection{Viscous Dissipation}

The viscous dissipation can be written as
\begin{equation}
F_{\rm diss}=\frac{1}{2}\eta \left( R\frac{d\Omega }{dR}\right)
^{2}  
\label{dissipation1}
\end{equation}
where $\eta $ is the vertically averaged dynamical viscosity and $\Omega $ is the angular
velocity of the matter in the disk. For a thin disk angular velocity is Keplerian 
$\Omega =\sqrt{GM/R^3}$
where $M$ is the mass of the neutron star.
For a Keplerian disk  \eqref{dissipation1} becomes
\begin{equation}
F_{\rm diss}=\frac{9}{8}\eta \frac{GM}{R^3}.
\label{dissipation2}
\end{equation}
The angular momentum balance in the steady state reads
\begin{equation}
\dot{J}_{\rm d}=\dot{M}_{\rm d} R^2\Omega +2\pi R^3 \eta \frac{d\Omega }{dR}
\label{torque}
\end{equation}
where $\dot{J}_{\rm d}$ is the flux of angular momentum (torque) and $\dot{M}_{\rm d}$ is
the mass flux (accretion rate) in the disk, both of which are integration
constants for a steady disk. The first term on the right hand side is the
material torque and the second term is the viscous torque. The torques acted by the star are the propeller torque due to  the displacement of  inflowing matter by the non-spherical magnetosphere \citep{ill75} and the magnetic torque due to coupling between the dipole field of the star and the toroidal field generated in the disk.
For a Keplerian disk the angular momentum balance equation can be organised as
\begin{equation}
\eta =\frac{1}{3\pi} \left( \dot{M}_{\rm d} - \frac{\dot{J}_{\rm d}}{\sqrt{GMR}} \right).
\label{nusigma1}
\end{equation}
In the general case one parametrizes
the torque as
$\dot{J}_{\rm d} \equiv \beta \dot{M}_{\rm d} R_{\rm in}^2\Omega (R_{\rm in})$
where $\beta $ is a dimensionless measure of the total torque applied on the star
by the disk. By using this Equation~\eqref{nusigma1} can be written as
$\eta =(\dot{M}_{\rm d}/3\pi) ( 1-\beta \sqrt{R_{\rm in}/R})$ and  
using this in Eq.~\eqref{dissipation2} one obtains
$F_{\rm diss}=(3GM\dot{M}_{\rm d}/8\pi R^3) ( 1-\beta \sqrt{R_{\rm in}/R})$
 which gives the dissipation by viscous stresses in the disk. For a purely viscous
disk (no irradiation) the temperature is found by 
$\sigma T^{4}=F_{\rm diss}$ and it means that $T\propto R^{-3/4}$ for $R \gg R_{\rm in}$.
Here the $\beta \simeq 1$ case corresponds to the accretion stage \citep{sha73} while $\beta \rightarrow -\infty$ with $\dot{M}_{\rm d} \rightarrow 0$ corresponds to the quiescent disk of SS77 describing the propeller stage. In this limiting case Equation~\eqref{nusigma1} reduces to
\begin{equation} 
\eta = \frac{-\dot{J}_{\rm d}}{3\pi \sqrt{GMR}}
\label{dissipation3}
\end{equation}
and using this equation, Eq.~\eqref{dissipation2} the flux is
\begin{equation}
F_{\rm diss}=\frac{3(-\dot{J}_{\rm d})\sqrt{GM}}{8\pi R^{7/2} }  \qquad \mbox{full propeller}
\label{dissipation4}
\end{equation}
yielding $T \propto R^{-7/8}$ (SS77) for a blackbody ($\sigma T^4 = F_{\rm diss}$). Hence the flux of a propelled disk is not only reduced compared to an accretion disk with no viscous torque at the inner boundary but also has a different temperature distribution that would lead to a spectrum with a different slope.

The luminosity of a thin layer of thickness $dR$ will be
$dL = 2(2\pi RdR) F_{\rm diss}$. The total luminosity of the Keplerian disk can be obtained by using Equation~\eqref{dissipation4} and integrating over the radial extend of the disk. Assuming the outer radius to be much greater than the inner radius of the disk one obtains 
\begin{equation}
L_{\rm K} = -\dot{J}_{\rm d} \Omega_{\rm K}(R_{\rm in}).
\label{L_K}
\end{equation}
It is likely that there is outflow of matter in the propeller stage.

\subsection{Irradiation of the disk}

Another contribution to the energy radiated by the disk is due to the
reprocessing of the X-ray luminosity $L_{\rm X}$ from a central source, the hot neutron star. 
We assume that hard X-ray and gamma-ray emissions from the magnetosphere are not beamed towards the disk as they could otherwise 
strongly ablate the disk. Since pulsations are observed, they are beamed
towards Earth and show low extinction.
Irradiation dominates viscous dissipation in the outer and cooler parts of the disk and explains the IR radiation.

\begin{figure}
\includegraphics[width=0.5\textwidth]{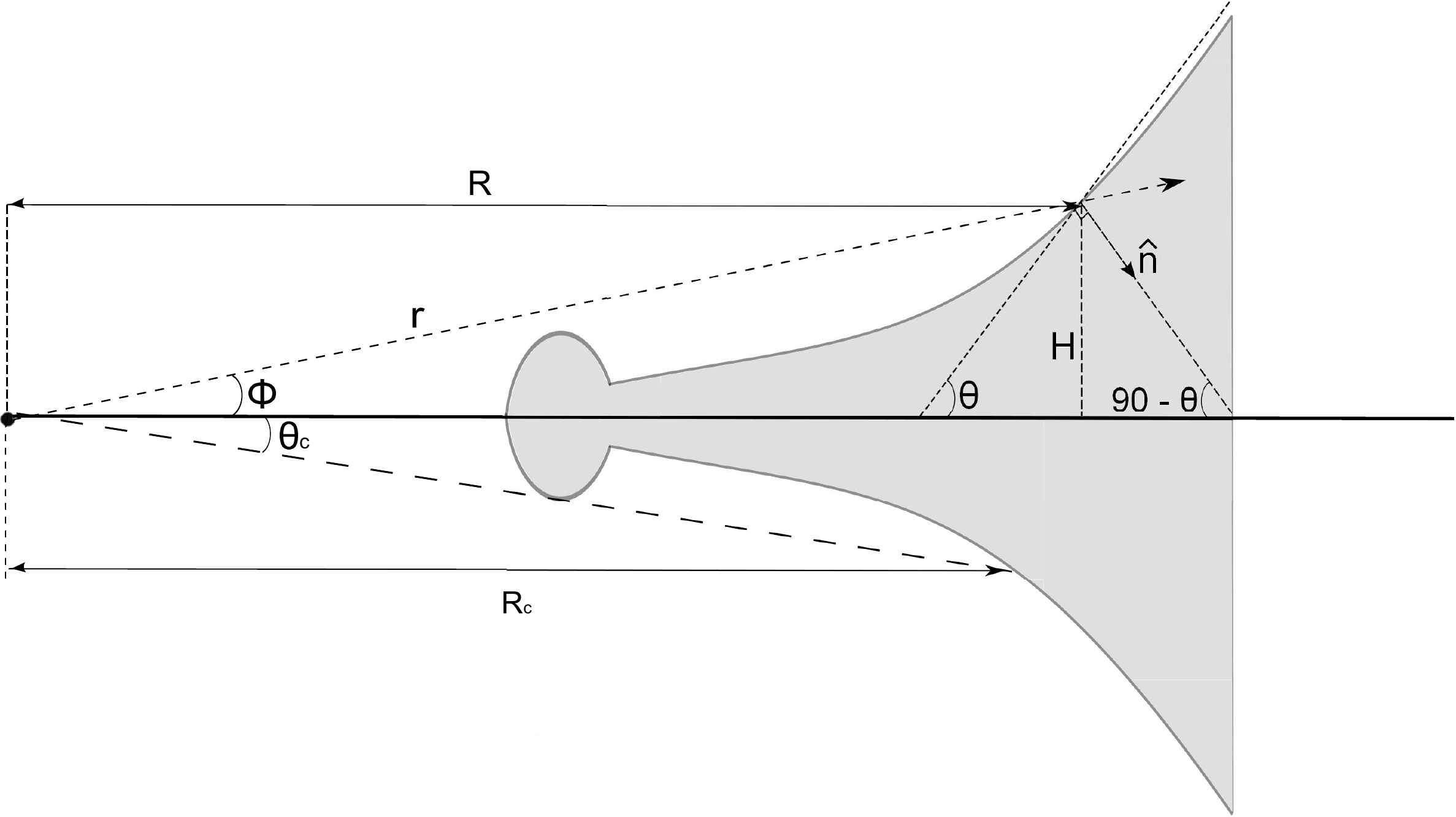}
\caption{Geometry of the star-disk system. The cylindrical radial distance from the compact object is shown with $R$ while $r$ is the radial distance. The scale-height of the disk is $H(R)$. The inner portion of the disk is relatively thick shadowing the part of the disk close to the disk mid-plane less than a critical angle $\theta_{\rm c}$. This defines a critical radius $R_{\rm c}$ beyond which the central object can irradiate the disk.}
\label{fig-geo}
\end{figure}

The flux of reprocessed emission of the irradiated disk is given by \citep{fra+02}:
\begin{equation}
F_{\rm irr}=\frac{L_{\rm X} (1-\alpha)}{4\pi R^2} \frac{H}{R} \left(\frac{d \ln H}{d \ln R } -1 \right).
\label{irr_central}
\end{equation}
where $H$ is the vertical scale-height of the disk, $\alpha$ is the albedo. The irradiated isothermal disk is flared as 
\begin{equation}
\frac{H}{R} = \left( \frac{L_{\rm X} f}{14\pi \sigma } \right)^{1/7} \left(\frac{k_{\rm B} }{\bar{\mu} m_{\rm p} G M}\right)^{4/7} R^{2/7}.
\label{HR2}
\end{equation}
so that $d \ln H/d \ln R = 9/7$ \citep{vrt+90,chi97} where $\bar{\mu}$ is the mean molecular weight and $f=1-\alpha$ also carries all uncertainties about the vertical structure. 
Solving for the temperature distribution by  $\sigma T^4 = F_{\rm irr}$, we obtain 
\begin{equation}
T=\left( \frac{L_{\rm X} f}{14\pi \sigma } \sqrt{\frac{k_B}{\bar{\mu} m_p G M}} \right)^{2/7} R^{-3/7}.
\label{irrad2}
\end{equation}

In the general case where both viscous dissipation and irradiation work, the
effective temperature at radial distance $R$ can be found from the
Stefan-Boltzmann's law
\begin{equation}
\sigma T^4=F_{\rm diss}+F_{\rm irr}.  \label{T_eff1}
\end{equation}
The spectrum of the disk is then the sum of
the blackbody contributions from all layers and is found by Equation (\ref{F_nu}).

\subsection{Boundary layer luminosity}

Using Equations (\ref{dissipation1}) and (\ref{torque}) the luminosity $dL = 2(2\pi RdR) F_{\rm diss}$ of a thin layer of thickness $dR$ can be written as
\begin{equation}
dL = (\dot{J}_{\rm d}-\dot{M}_{\rm d} R^2 \Omega) d\Omega.
\label{dL}
\end{equation}
Using $\dot{J}_{\rm d} \equiv \beta \dot{M}_{\rm d} R_{\rm in}^2\Omega (R_{\rm in})$ and integrating the above for a narrow $b \ll R_{\rm in}$ boundary layer where $b$ is the width, we find
\begin{equation}
L_{\rm BL} = \frac{GM\dot{M}_{\rm d}}{2R_{\rm in}} (2\beta - 1 -\omega_{\ast})(1-\omega_{\ast})
\label{L_bl}
\end{equation}
where $\omega_{\ast} \equiv \Omega_{\ast}/\Omega_{\rm K}(R_{\rm in})$
is the fastness parameter.
For the accretion stage $\beta \simeq 1$ and Equation~(\ref{L_bl}) reduces to the well known result
\begin{equation}
L_{\rm BL} = \frac{GM\dot{M}_{\rm d}}{2R_{\rm in}} (1-\omega_{\ast})^2, \qquad {\rm if}\,\,  \omega_{\ast}<1
\end{equation}
\citep[see][p.\ 154]{fra+02}. 
At the extreme limit of full propeller ($\dot{M}=0$), 
Equation~(\ref{dL}) becomes $dL=\dot{J}_{\rm d}d\Omega$
which can be integrated to give $L_{\rm BL} = -\dot{J}_{\rm d} (\Omega_{\ast} - \Omega_{\rm K}(R_{\rm in}))$
so that
\begin{equation}
L_{\rm BL} = -\dot{J}_{\rm d} \Omega_{\rm K}(R_{\rm in}) (\omega_{\ast}-1), \qquad {\rm if}\,\, \omega_{\ast} > 1
\label{L_BL}
\end{equation}
which, by using Equation~\eqref{L_K}, implies $L_{\rm BL} / L_{\rm K} = \omega_{\ast}-1$ in the propeller regime.

The effective temperature of the boundary layer will be found by 
$L_{\rm BL} = 2 (2\pi R_{\rm in}) b \sigma T_{\rm BL}^4$.
Defining $\epsilon \equiv b/R_{\rm in}$ the boundary layer temperature becomes
\begin{equation}
T_{\rm BL}  = \left( \frac{L_{\rm BL}}{4\pi R_{\rm in}^2  \sigma \epsilon} \right)^{1/4}.
\end{equation}
We have found $\epsilon \sim 0.01$ by fitting the spectrum (see \S~\ref{sec-spectrum}).
The intensity will be
\begin{equation}
I_{\nu}=\frac{2h\nu^3/c^2}{{\rm e}^{h\nu /k_{\rm B}T_{\rm BL}}-1}
\label{bl_bb}
\end{equation}
so that
\begin{equation}
F_{\nu, {\rm BL}} = \frac{4\pi h \cos i}{c^2} \left(\frac{R_{\rm in}}{d} \right)^2 \frac{\epsilon \nu^3}{e^{h\nu/k_B T_{\rm BL}} - 1}
\end{equation}

In addition to the viscous dissipation there is Ohmic dissipation in the disk. 
We neglect the latter as \citet{lov+95} show that Ohmic dissipation is much smaller 
than the viscous dissipation.

\subsection{Puffed up inner rim of the disk}
\label{puffed}

In order to obtain a good fit to the spectrum we needed shadowing of some part of the disk by 
the geometrically thick inner parts of the disk (see Figure~\ref{fig-geo}).
This is similar to the procedure used in modelling the spectrum of 
protoplanetary disks---the puffed up inner rim \citep{nat+01,dul+01}. 
The origin of the puffed up inner rim in the context of protoplanetary disks is the heating by the central source \citep{dul+01}. In the context of propeller disks around compact objects the puffed up inner rim could well be the boundary layer which is expected to be geometrically thick due to its own luminosity.

The geometrically thick boundary layer shades a substantial fraction of the inner disk up to $R_{\rm c}$. This corresponds to a critical angle $\Phi_{\rm c}$ such that $\tan \Phi_{\rm c} = (H/R)_{\rm c}$ and could be found by Equation (\ref{HR2}).

\section{Modeling the spectrum} 
\label{sec-spectrum}

We have modelled the unpulsed spectrum of the Vela pulsar assuming it arises from the disk. 
We have employed the semi-amplitude of modulation
\begin{equation}
A = \frac{\sum_i |{\rm CR}_i-{\rm CR}_{\rm mean}|}{\sum_i {\rm CR}_i},
\label{semi_amplitude}
\end{equation}
rather than the pulsed-fraction $PF={\rm CR}_{\max}-{\rm CR}_{\min}/({\rm CR}_{\max}+{\rm CR}_{\min})$,
as an appropriate descriptor of the pulsed emission for complex shape light curves. Here ${\rm CR}$ is the count-rate. For the optical band we have analysed the pulse profiles given in \citet{man+80} and found $A=0.34$; for the UV band we have analysed \citet{rom+05} and determined $A=0.51$ for the near UV and $A=0.45$ for the far UV. In the absence of pulse profiles in the IR, we have assumed the semi-amplitude of modulation in the IR is the same as that of the optical. The observational flux values are compiled from the literature as follows: JH and mid-infrared data are taken from \citet{shi+03}, \citep{dan+11} and \citep{zyu+13}, and UV and optical data are taken from \citet{rom+05} and \citet{mig+07opt}. 
We have re-reduced the J, H, K and HST data, except for F195W
and G140 as described in \S \ref{sec-photometry}. In fitting the spectrum we have employed our measurements where available, but we have seen that the parameters do not differ significantly. The 68\% and 95\% HPD intervals obtained by MCMC fitting show the combined range.

\begin{figure}
\includegraphics[width=0.5\textwidth]{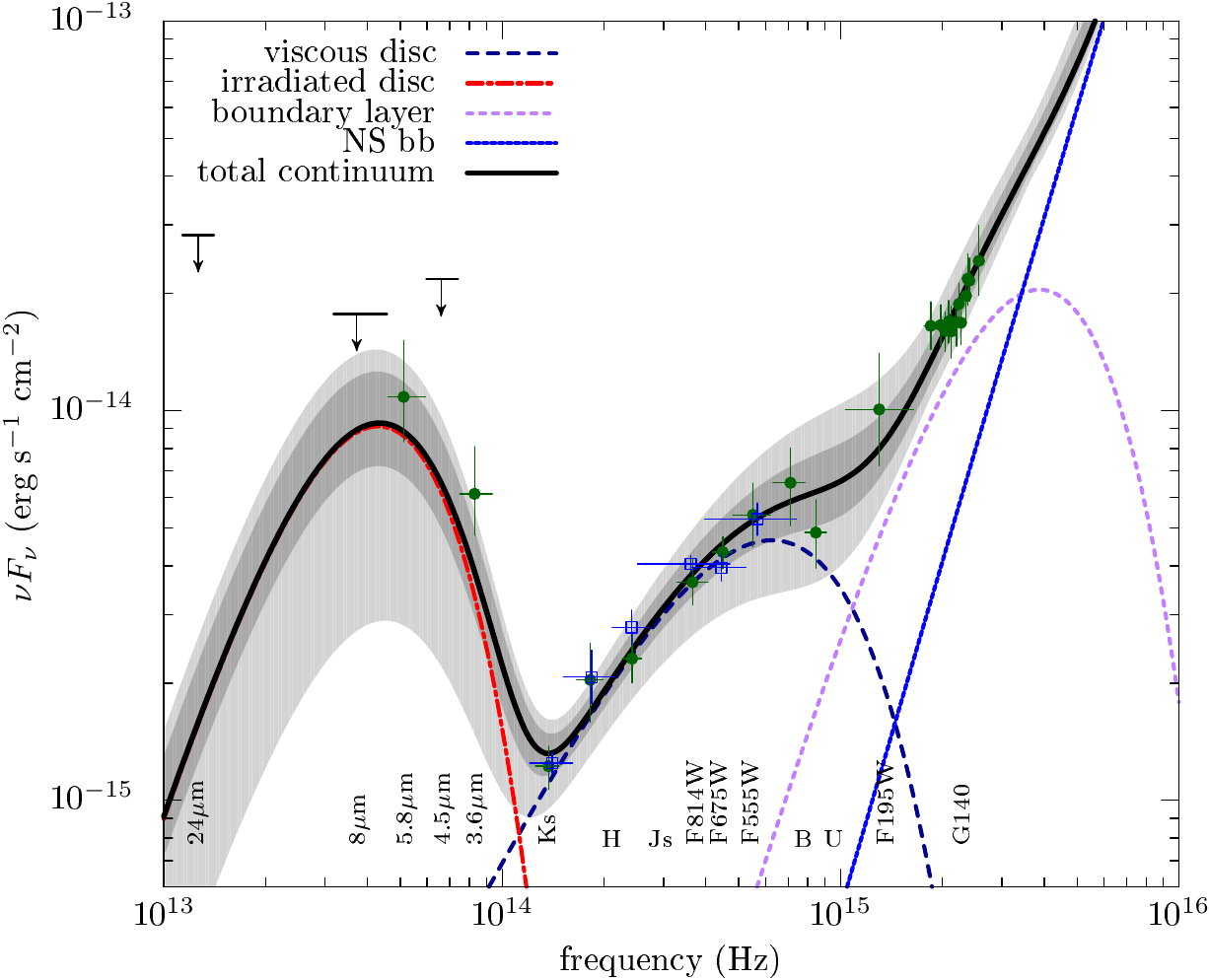}
\caption{The spectrum of the Vela pulsar fitted with the model described in \S \ref{sec-model} including viscous dissipation (long dashed line--colored blue in the electronic version), irradiation (dashed and dotted line --colored red in the electronic version) and boundary layer emission appropriate for propellers (short dashed line--colored purple in the electronic version). The solid line shows the total theoretical spectrum. The dark and light grey shaded regions are the 68\% and 95\% HPD intervals obtained by MCMC fitting (6 chains, each 250\,000 iterations), including observed data with upper limits as censored ones in the likelihood. The observational data (solid circles---colored green in the electronic version) are taken from \citet{dan+11} \citep[see also][]{shi+03,rom+05,mig+07opt}; J, H, K and HST data, except for F195W
and G140, were re-reduced by us (see \S \ref{sec-photometry}) and are shown with hollow squares (colored blue in the electronic version). Data plotted and modelled are the
unpulsed contribution. For the MCMC fitting integrated spectra  of the total theoretical spectra over each band are used.}
\label{fig-vela}
\end{figure}

We fit the infrared, optical and UV spectrum of Vela with the model described in \S \ref{sec-model} (see Figure \ref{fig-vela}) by employing a Markov Chain Monte-Carlo (MCMC) method in which the 6 fit parameters are the inner and outer radii of the disk, the boundary layer luminosity, the width of the boundary layer ($b$), the critical radius beyond which the disk is not shaded ($R_{\rm c}$) and the radius of the neutron star's cold black body component, $R_{\rm bb}$ in units of 10~km, $f_{\rm bb}$. The outer part of the disk is irradiated by the thermal X-ray radiation of the neutron star. The viscous dissipation in the disk addresses the optical part of the spectrum ( H, Js and F814W bands) while the boundary layer luminosity of the disk dominates in the UV band where an extrapolation of the cold component of the black body radiation from the surface of the neutron star also contributes. 
We find the inner radius of the disk by spectral fitting to be $R_{\rm in}= 0.88^{+0.04}_{-0.06} \times 10^{8}$~cm well inside the light cylinder radius ($R_{\rm in} \simeq 0.21 R_{\rm L}$) and 2.63 times the corotation radius $R_{\rm co} = (GM/\Omega_{\ast}^2)^{1/3}=0.335\times 10^8$~cm, corresponding to a 
fastness parameter of $\omega_{\ast} =(R_{\rm in}/R_{\rm co})^{3/2} \simeq 4.3$, placing this object well in the propeller stage \citep{shv70,ill75}. A fraction of the spin-down luminosity of the pulsar is revealed in the disk-magnetosphere boundary \citep{eks+05}. We find the boundary layer luminosity to be $2.39^{+0.18}_{-0.24} \times 10^{29}$~erg~s$^{-1}$, very small compared the spin-down luminosity. In order to save the disk model one is forced to assume that most of the energy transferred to the disk by the star is taken up by the kinetic energy of the outflow and the expansion of the disk. 
In the case of an outflow the disk torque $\dot{J}_{\rm d}$ in Eqns.~\eqref{L_K} and \eqref{L_BL} is to be replaced by the effective torque $\dot{J}'=\dot{J}_{\rm d} + \dot{M}_{\rm out}R_{\rm in}^2 \Omega_\ast$. But matter escape by the outflow would require transport of similar amount of matter within the disk and this would lead to extra viscous dissipation unless the disk is advection dominated \citep{nar94,ich77}. Investigation of such disks with outflow is beyond the scope of the present paper.
The width of the boundary layer is fit to be $b = 7.56^{+1.14}_{-1.44}\times 10^5$~cm ($\epsilon=b/R_{\rm in}\simeq 1\%$). 
This then gives the effective temperature of the boundary layer to be $T_{\rm bl} =4.7\times 10^4$~K.
We have found, $f_{\rm bb} \equiv R_{\rm bb}/10~{\rm km} = 0.454^{+0.0395}_{-0.0025}$ consistent with the $R_{\rm bb}=5.06^{+0.42}_{-0.28}~{\rm km}$ of \citet{man+07}.

The outer parts of the disk dominate the infrared spectra. The emission from this region of the disk is due to the reprocessing of the X-ray luminosity $L_{\rm X}=3.1 \times 10^{32}$~erg~s$^{-1}$ of the central object. We used the temperature distribution given in Equation (\ref{irrad2}) as
\begin{equation}
T(R_{10}) = 8.36 \times 10^2 \, {\rm K} \left(\frac{f}{0.5}\right)^{2/7} 
L_{32}^{2/7} R_{10}^{-3/7}
\end{equation}
\citep{vrt+90} where $L_{32} = L_{\rm X}/10^{32}\, {\rm erg\, s^{-1}}$ and $R_{10}=R/10^{10}\, {\rm cm}$. The uncertainty factor $f$ is taken to be 0.5 in \citet{vrt+90}.
We find, from our spectral fit, the outer radius of the disk to be
$R_{\rm out}= 8.21^{+1.05}_{0.63} \times 10^{10}$~cm. 

In order to obtain a good fit we have also needed the shadowing of the inner parts of the disk due to a geometrically thick boundary layer (see Figure~\ref{fig-geo} as described in \S \ref{puffed}). We fit this critical radius beyond which the boundary layer can not screen the disk to be $R_{\rm c} = 4.80^{+0.38}_{-0.23} \times 10^{10}$ cm, more than half of the disk. From this we find $\tan \Phi_{\rm c}=(H/R)_{\rm c}= 0.0045$ which means that such substantial shading of the disk is achieved by a boundary layer subtending an angle of 
0.26 degrees only.

\section{X-ray data analysis}
\label{sec-X-ray}
 Naturally rises the question about the possibility of observing some spectral features owing 
to the reprocessed X-ray radiation by the fallback disk.
Indeed, using fitted parameters of the fallback disk model and ionizing central source characteristics, we employed the XSTAR\footnote{\url{http://heasarc.nasa.gov/lheasoft/xstar/xstar.html}} code in order to assess the equivalent width
(EW) of OVII ions, already observed in a number of astrophysical cases
and attributed to the circumstellar environment  \citep[e.g.][]{ham+09}. For the ionizing source we adopted a two blackbody spectrum with temperatures of 86 and 173 eV  and a luminosity of 
$L_{\rm bol} \sim 3.1 \times 10^{32}$~erg~s$^{-1}$ of the neutron star \citep{man+07}, a hydrogen density of $n_{\rm H}\sim 10^8-10^{10}$ cm$^{-3}$ allocated in the outer part 
of the fallback disk, having a tiny covering fraction and extended up to $5\times 10^{11}$~cm.
This model will provide the ionization parameter of $\log \xi \sim 1-3$ ($\xi=L_{\rm X}/n_{\rm H} r^2$). 
Simulations resulted in a $0.5-1.3$~eV EW of OVII ions which can be readily
measured in XMM-Newton RGS high resolution spectra.

\subsection{Data reduction}

The {\it RGS} data were reduced using the {\it rgsproc} task. In all cases only events from time intervals 
with a count rate below 0.2~cts/s (determined for chip number nine of both {\it RGS} using timebinsize=100) are analyzed. 
The individual {\it RGS} spectra were added using the {\it rgscombine} task that also produces the combined response 
matrices and background spectra of the co-added source spectra for RGS1 and RGS2.

Since Vela exhibits diffuse emission at X-ray energies due to its pulsar wind nebula (PWN) and therefore 
does not appear as a point source \citep{pav+03}, we also used the task {\it rgsrmfgen} by supplying an intensity
 profile of the source along the direction of the dispersion axis for each observation individually. This reduces a 
broadening of possible spectral features, if the emission does not originate from a perfect point source.

However, measuring the intensity profile along the direction of the dispersion axis is only feasible for the last four 
observations (0510390301, 0510390501, 0510390601, 0510390701), since they are performed in {\it EPIC-pn} large window mode. 
In all other cases, the observations are performed in small window mode for {\it EPIC-pn} and both {\it EPIC-MOS}. 
We obtained the intensity profile (including the point source and the extended emission) from a projection 
region with {\it DS9} version 7.0.2 from each {\it EPIC-pn} (large window mode) image, that produces an ascii file. This file lists the intensity of each pixel value that we converted (using a small {\it Matlab~R2012a} routine) into starting angle and stop angle (accounting for the roll angle of the telescope) in units of arcmin (depending on image resolution). The modified file can be provided to {\it rgsproc} via the command {\it angdistset}.

\subsection{Spectral analysis of high resolution spectra}
\label{highres}

Adding all available {\it RGS} spectra sums up to a total net exposure time of 737~ks ({\it RGS1}) and 734~ks ({\it RGS2}) with 640,000 and 688,000 source photons, respectively. We grouped the source photons of both {\it RGS} to 2000 counts per bin using the command {\it grppha} to even further improve statistics.

The co-added {\it RGS1} and {\it RGS2} spectra were fitted together in one session, i.e.\ the model parameters (with the {\it Xspec} model {\it phabs(bbodyrad+bbodyrad+pow)}, see \citet{man+07} of the {\it RGS2} spectrum are fixed to the fit values of the corresponding model parameters of the {\it RGS1} spectrum.

Technically, the fit quality is low ($\mathrm{\chi^2_{red}=1.89}$). This low quality is not caused by an unsuitable model, but rather by calibration problems due to numerous chip-gaps and dead pixels (see the {\it RGS} user manual) in the {\it RGS}.

The fit residuals exhibit a possible absorption feature at 0.57~keV. Therefore, we included {\it gabs} as further component in the fit model to account for this feature. The {\it gabs} model yields line energy, optical depth and sigma of the line. However, the equivalent width of a line is a more reliable parameter, in particular on estimating the line significance. Since {\it gabs} is not an additive model, {\it Xspec} does not provide a direct command to calculate the equivalent width. Therefore, we calculated the equivalent width by providing the line parameters derived from the {\it gabs} model (assuming a Gaussian distribution according to their errors) to a {\it Matlab~R2012a} routine, running a Monte-Carlo-Simulation ($\mathrm{50^3}$ parameter sets). The output equivalent widths are Gaussian distributed, delivering the corresponding confidence interval.

Indeed, the fit improves ($\mathrm{\chi^2_{red}=1.85}$), yielding an equivalent width of $\mathrm{EW=-2.0\pm1.2~eV}$ (90\% confidence) for the previously unknown absorption feature.
 
Since the emission of Vela appears extended, we furthermore co-added only those four observations, where it was possible to measure the intensity profile along the {\it RGS} dispersion axes. The resulting {\it RGS} spectra sums up to a total net exposure time of 413~ks (both {\it RGS}) with 349,000 and 376,000 source photons, respectively.

Analysing the data in exactly the same manner yields $\mathrm{EW=-4.2\pm 1.3}$ (again $\mathrm{\chi^2_{red}\approx2}$, even if the absorption feature is included in the fit model). Although the second spectra include much less photons compared to the previous methods, the line appears to be more significant (2.8$\sigma$ versus 5.2$\sigma$). This effect can be explained by assuming a contribution from an extended emission region, such as a hot disk, around the Vela pulsar. In the first methods, these parts of the line would smear out and thus would not contribute to the spectrum. This implies that the line significance changes by accounting for possible contributions from an extended emission region. According to the {\it RGS} manual, the size of the emission region must be of the order of tens of arcsec or even some arcmin, i.e.\ $\mathrm{10^3-10^4~AU}$ (if scaled to Vela's distance) to cause such effects in the {\it RGS} spectra. 
Therefore, we consider the jet and/or the PWN as source of the absorption feature from the extended emission and an overabundance of material (related to the ISM) from the point source, that is a strong evidence for the presence of a disk around Vela.

\begin{figure}
\includegraphics[width=0.48\textwidth]{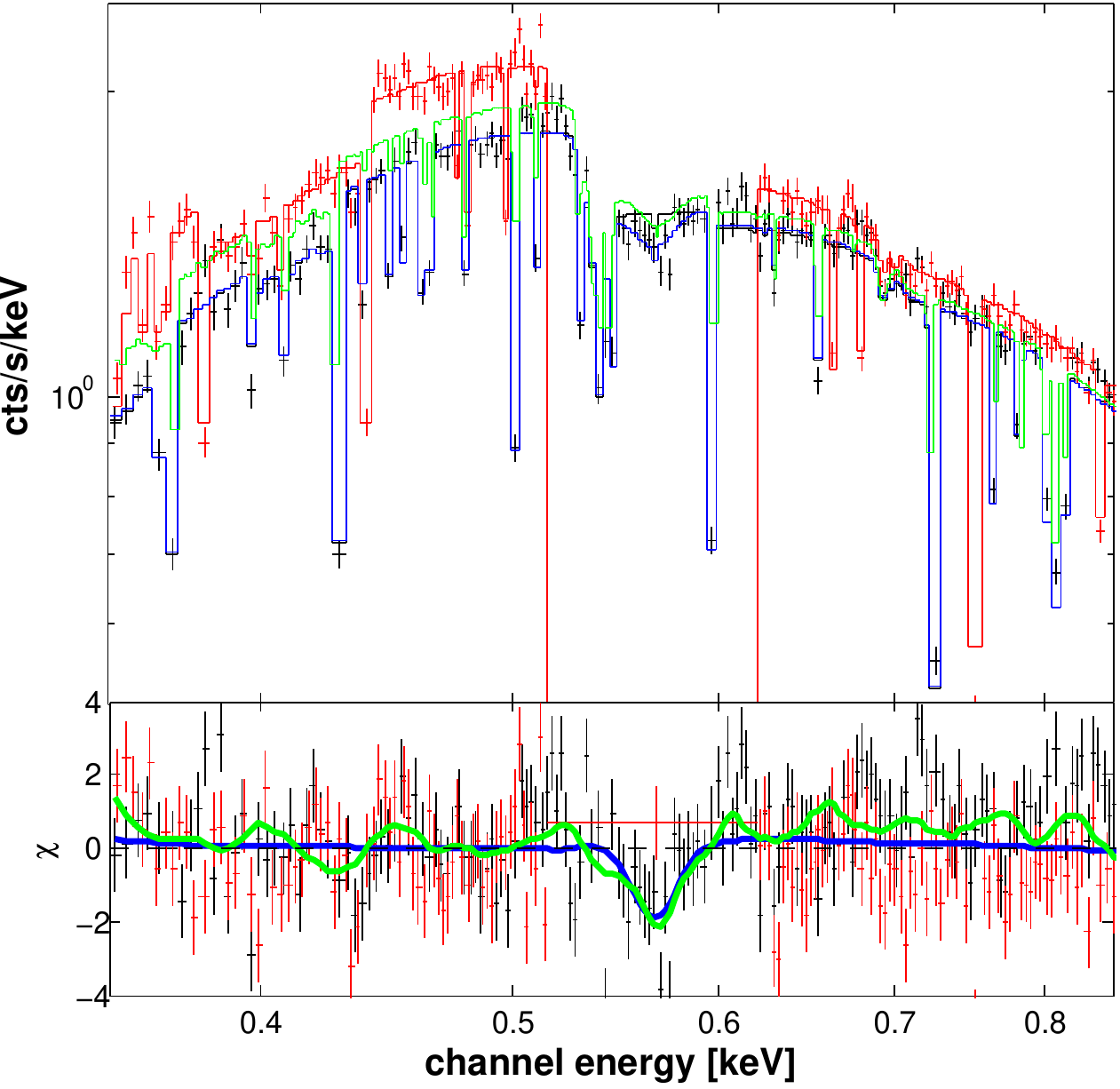}
\caption{Co-added {\it RGS1} (black) and {\it RGS2} (red) spectra from those four Vela observations, where the intensity profile of the source along the {\it RGS} dispersion axes could be measured. The blue solid lines show the fit model including a narrow absorption feature at 0.57~keV derived from these spectra. As comparison we also show the fit model including the feature derived from co-adding all {\it RGS} data, regardless of the intensity profile (green solid line). In all cases we used the model {\it TBabs} \citep{wil+00} to account for the contribution (absorption/extinction) of the ISM.}
\label{rgs}
\end{figure}

Furthermore, we calculated the equivalent width by fitting the {\it RGS} data using an alternative model ({\it TBabs}) for the ISM. Again, the line appears stronger, if the emission of an extended region is taken into account. Our results of the line parameters are shown in \autoref{fitvalRGS}, the spectra are shown in \autoref{rgs}.

The $\mathrm{K_{\alpha}}$ triplet of OVII contains the resonance line (all line energies at rest) at 573.95~eV, the intercombination line at 568.63~eV and the forbidden line at 560.99~eV  (\citet{Hirata} and \citet{nist}). In all cases, the measured line energies (\autoref{rgs}) are consistent with the line energy of the intercombination line, but not with the other line energies. However, the intercombination line should not appear in absorption. This (still unsolved) enigma does also appear in the case of the isolated X-ray pulsar RX\,J0720.4$-$3125 \citep{hoh+12,ham+09}, whereas the absorption feature found in the {\it RGS} spectra of RX\,J1605.3+3249 is consistent with the resonance line \citep{hoh+12,ker+04}. We stress, however, that the errors of the line energies are relatively large in spectra fitting with $\mathrm{\chi^2_{red}\approx2}$ and that the spectral resolution of {\it RGS} amounts 1--2~eV at 0.5~keV. Thus, an exact line identification requires further observations.

According to the $\mathrm{N_H}$ values \citep{man+07} of $\mathrm{N_H=3-5.7\times10^{20}/cm^2}$ and an oxygen/hydrogen ratio of  $\mathrm{n_O/n_H=4\times10^{-4}}$ \citep{and89} in the solar vicinity, 
the expected column density of OVII (ISM only, all oxygen ionized to OVII) is $\mathrm{N_{OVII}\approx1.2-2.3\times10^{17}/cm^2}$ in the line of sight. 
Note that in order to provide significant amount of OVII ions an unusual hot component 
of the ISM of temperature of 1.2MK must present in the direction of Vela pulsar. 
On the other hand, the equivalent widths of the absorption feature from both methods (\autoref{fitvalRGS}) yield \citep{fut+04} $\mathrm{N_{OVII}\approx 10^{16}-5\times10^{19}/cm^2}$ (left column in \autoref{fitvalRGS}) and $\mathrm{N_{OVII}\approx 10^{17}-5\times10^{20}/cm^2}$ (right column in \autoref{fitvalRGS}), respectively. Hence, even if all oxygen is ionized to OVII (that is not possible, since then already there should be OVIII and some OVI is still left in case of equilibrium) and the velocity dispersion of the particles 
is above 420~km/s \citep[][Fig.~2 therein]{fut+04}, the equivalent widths are hard to reconcile with being caused by the contributions from the ISM only. 

Thus, the co-added  {\it RGS} spectra of Vela were fitted with two different models for the 
ISM and taking the extended emission of Vela into account. The resulting {\it RGS} spectra, shown 
in Figure~\ref{rgs} show strong evidence for the presence of a narrow absorption feature at 0.57~keV. 
This feature can be interpreted as the $\mathrm{K_{\alpha}}$ line of He-like oxygen (OVII) and thus 
points to a significant over-abundance of these ions in the line of sight. By measuring the line strength, 
the required amounts of material would exceed these as expected from the ISM. 
Therefore, the narrow absorption feature at 0.57~keV originate partly from a putative disk around the Vela pulsar 
and partly from the PWN, as shown by our methods of data analysis.

\section{Discussion}
\label{sec-discuss}

\begin{figure*}
\includegraphics[width=0.33\textwidth]{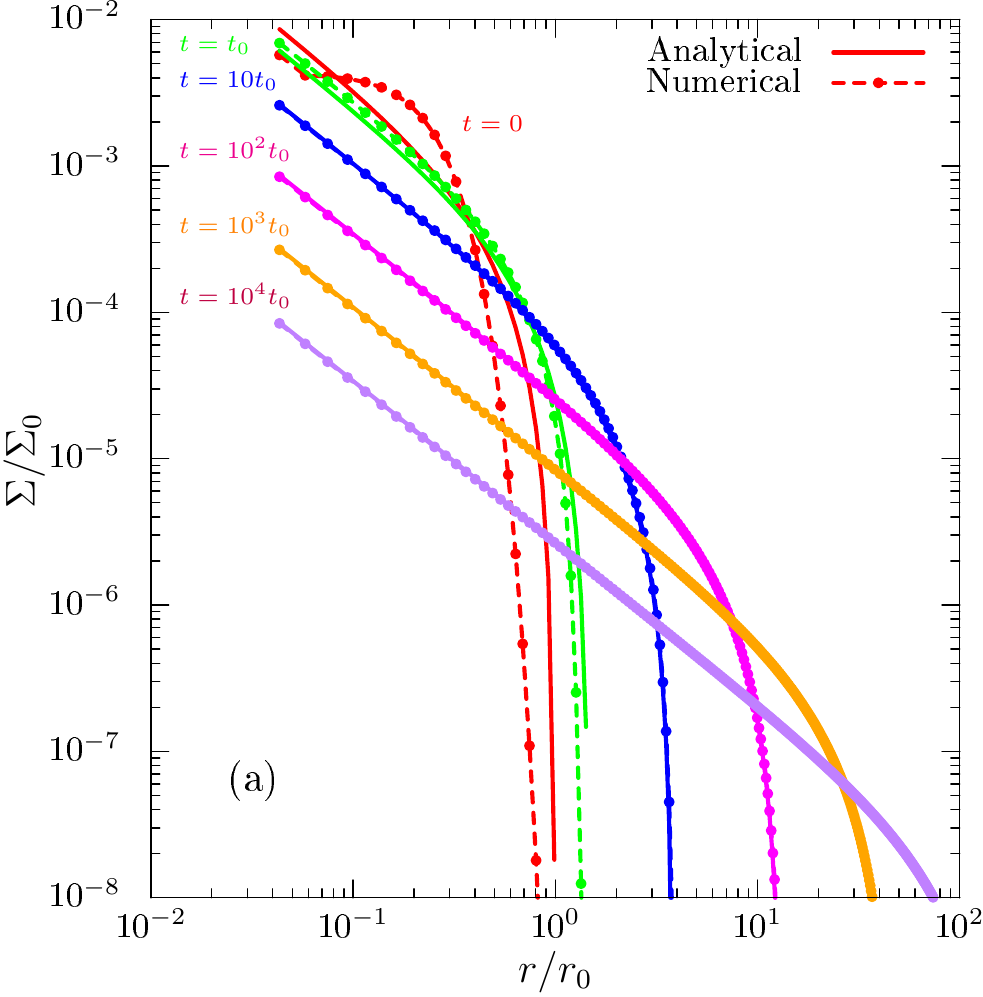}
\includegraphics[width=0.33\textwidth]{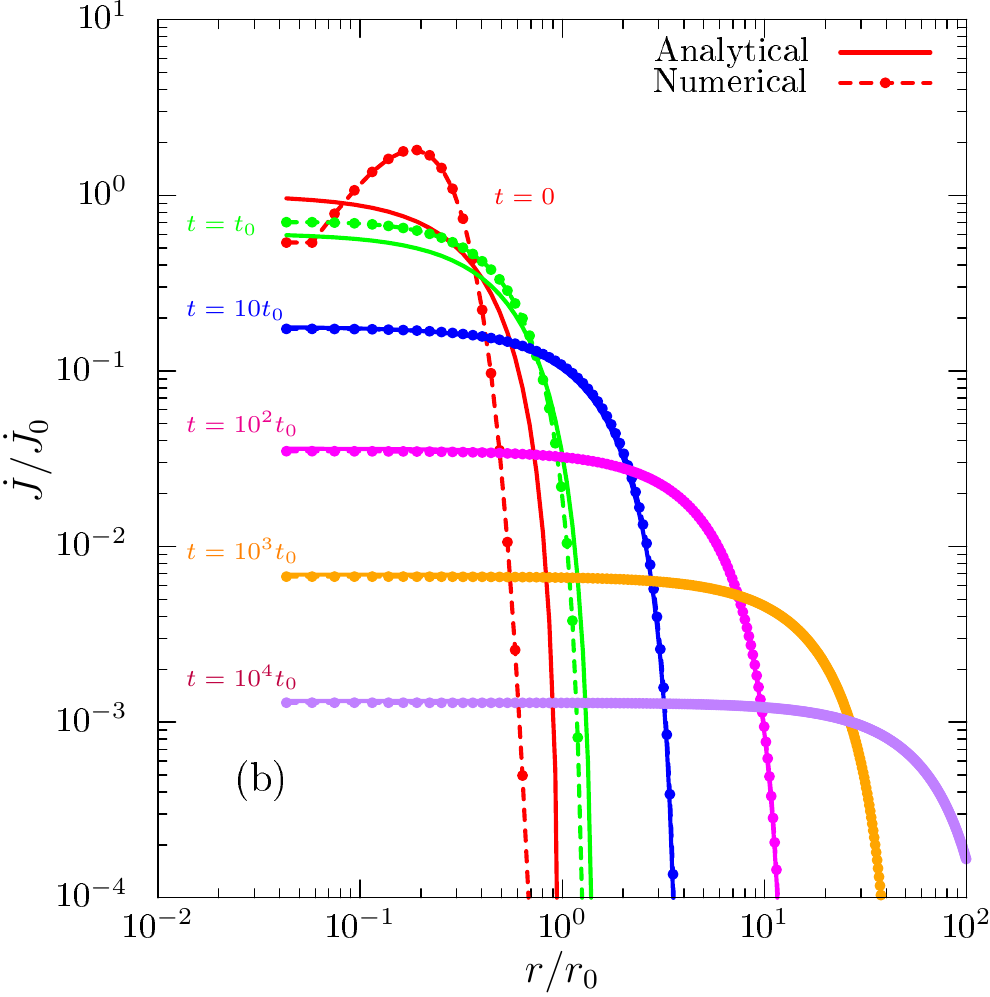}
\includegraphics[width=0.33\textwidth]{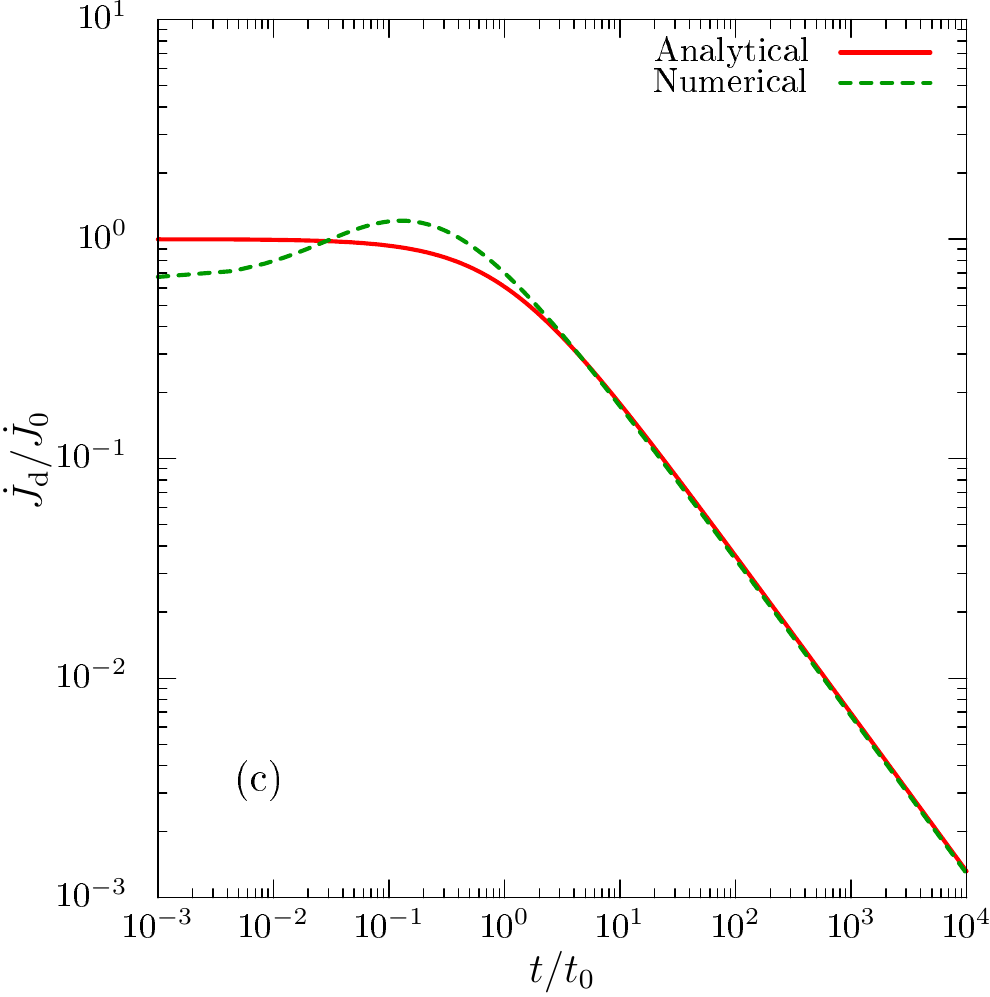}
\caption{The radial distribution of the surface density (panel a), the radial distribution of the viscous torque within the disk (panel b) and the time evolution of the viscous torque at the inner boundary of the disk (panel c) as obtained from numerical (dashed lines with points) and analytical (solid lines) solution of Eqn.~\eqref{diff1} with boundary condition $\dot{M}_{\rm d}=0$ assuming bound-free opacity prevails throughout the disk at all times. Initial distribution of the surface mass density $\Sigma$ in the numerical solution is chosen as a Gaussian with a peak near the inner boundary (colored in red in the electronic edition) and as such it does not match the analytical solution given in Eqn.~\eqref{sol2}. The mass and angular momentum under this curve is calculated, and used for determining $r_0$, $\Sigma_0$, $\nu_0$ and $t_0$ via Eqns.~\eqref{R_0} to \eqref{t_0} of the analytical solution, respectively. It is seen that the numerical solution approaches the analytical solution in about 10 viscous-timescales ($t_0$) during which the initial conditions are forgotten. On panel c the solid  line (colored in red in the electronic version) stands for the analytical solution $\dot{J}_{\rm d}/\dot{J}_0 =  \left(1 + t/t_0 \right)^{-18/25}$ from Eqn.~\eqref{Jdot_d_analytical}. Dashed line
(colored in green in the electronic edition) stands for the numerical solution. }
\label{fig-torque}
\end{figure*}

\subsection{Torque evolution}
\label{sec-torque}
 The evolution of a thin disk, such as the one we suggest to exist around the Vela pulsar, is described by a diffusion equation for the surface mass density, $\Sigma$ (see Eqn.\eqref{diff1} in Appendix). The analytical solution corresponding to the propeller stage i.e.\ with the boundary condition $\dot{M}_{\rm d}=0$ is given by \citet{pri74} \citep[see also][]{pri91}. In Appendix~\ref{m_0_j_0} this solution is described in terms of the initial mass and angular momentum of the disk. According to this solution a positive torque is applied to the disk at the inner radius which evolves with a power-law given in Equation~\eqref{Jdot_d_analytical}. The torque acted by the disk on the neutron star is then the negative of this and, for a bound-free opacity dominated disk, evolves as 
\begin{equation}
\dot{J}_{\rm d} = -\dot{J}_0 \left(1 + \frac{t}{t_0} \right)^{-18/25}
\label{viscous_torque}
\end{equation}
where $\dot{J}_0=7J_0/25t_0$ is the initial torque in terms of the initial angular momentum $J_0$ and the viscous time-scale $t_0=4R_0^2/3\nu_0$. The latter depends on the initial outer radius of the disk $R_0$ and turbulent viscosity scale $\nu_0=CR_0^{15/14}\Sigma_0^{3/7}$ where $C$ is a constant determined from the disk equations (see Eqn.\eqref{CC} and \citet{can+90}). 
In these equations $R_0=4.95 (J_0/M_0)^2/GM$ (see Eqn.~\eqref{R_0}) and 
$\Sigma_0 = M_0/2\pi R_0^2 \gamma_m$  (see Eqn.~\eqref{M_0}) with $\gamma_{\rm m}=1.3326\times 10^{-4}$. The evolution of the surface mass density, $\Sigma$  is shown in Figure~\ref{fig-torque} (panel a). We start the system from a Gaussian distribution for $\Sigma$ and determine the corresponding initial mass and angular momentum  of the disk which then determine $R_0$, $\Sigma_0$, $\nu_0$, and $t_0$ as given above. We see that after about 10 viscous time-scales the initial conditions are forgotten and the numerical solution matches the analytical solution. In panel (b) we show the viscous torque $\dot{J}=2\pi R^3 \eta \Omega'$  within the disk in units of $\dot{J}_0$. We note that after the self-similar stage is achieved the torque throughout the disk is radially constant up to the outer radius. In the steady state (infinite outer radius) the torque is a constant of motion. The self-similar stage is thus a quasi-equilibrium stage which resembles the steady state at each snapshot in time except for the outer radius being finite. In panel (c) we show the evolution of the viscous torque at the inner boundary. We emphasize that no fitting algorithm is applied in the figures apart from employing the same amount of initial mass and angular momentum in numerical and analytical solutions.

\subsection{Star-disk coupling}
\label{sec-couple}

Wherever the magnetic field of the star threads the disk it is twisted azimuthally due to the differential rotation between the magnetosphere and the disk. The growth of the toroidal field in the magnetically threaded domain will be limited by reconnection and turbulent diffusion \citep[see e.g.][]{wan95}. The stellar magnetic field can not thread the whole disk because then the large toroidal fields, generated in the regions where the local velocity difference $\Delta v = R (\Omega_{\ast}-\Omega)$ is large, would distrupt the disk \citep{wan87}. 
The field lines will favor an open topology \citep{aly90,uzd02} whenever the differential rotation angle exceeds about one radian \citep[see][for a review]{uzd04} and may remain open \citep{lov+95} or display a quasi-periodic opening and reconnection episodes \citep{bal94,goo+97}.

As being in a strong propeller regime the differential rotation between the disk and the star is very large throughout the disk except for a narrow region of width $\Delta R$ within the boundary layer where the angular velocity profile in the disk matches the angular velocity of the star. We assume the coupling to occur between $R_{\rm in}$  and $R_{\rm in}+\Delta R$ and define $\epsilon \equiv \Delta R / R_{\rm in} \ll 1$.

The torque exerted by the magnetic field lines threading the disk of radial extent $dR$ will be $d\dot{J}_{\rm m} = -R^2 B_{\phi}^{+} B_z dR$
where $B_{\phi}^{+}$ is the toroidal field evaluated at the disk surface. 
Assuming a dipole structure $B_z = - \mu/R^3$
and defining the twist (pitch angle) of the field lines
\begin{equation}
\gamma (R) \equiv \frac{B_{\phi}^{+}}{B_z}
\end{equation}
we can write
\begin{equation}
d\dot{J}_{\rm m} = -\gamma R^2  B_z^2 dR = -\mu^2 \frac{\gamma(R) dR}{R^4}.
\label{mag_torque}
\end{equation}
Assuming $\gamma$ changes linearly from zero at $R_{\rm in}$ to $\gamma_{\rm c}$ at $R_{\rm in} +\Delta R$ we find $\gamma = \gamma_{\rm c} R/(R_{\rm in} +\Delta R)$ and integrating Eqn.\eqref{mag_torque} we find
\begin{equation}
\dot{J}_{\rm m} = -\gamma_{\rm c} \epsilon \frac{\mu^2}{R_{\rm in}^3}.
\label{mag_torque2}
\end{equation}
This is an equation for $\epsilon$ rather than the magnetic torque (see the next section).

\subsection{Inner radius of the disk}
\label{sec-innerradius}

Our fit to the spectrum of the putative disk around the Vela pulsar gave a very low mass flux 
$\dot{M} \simeq 10^{10}$~g~s$^{-1}$ appropriate with the system being in the propeller stage. This corresponds to a very low ram pressure $\rho v^2$ and can not balance, within the light cylinder radius, the magnetic pressure $P_{\rm mag}=\mu^2/8\pi R^6$ of the dipole field of the neutron star where $\mu$ is the magnetic moment of the star. Indeed the Alfv{\' e}n radius
$R_{\rm A} = \mu^{4/7}(\sqrt{2GM}\dot{M})^{-2/7}$
\citep{dav73}, as determined from the equilibrium of magnetic pressure of the star with the ram pressure  of the accretion flow, is not appropriate for estimating the inner radius of fallback disks in the propeller regime. 

In the propeller stage matter builds up near the magnetospheric boundary \citep[see fig.\ 1 in ][]{rap+04} and the gas pressure (as well as viscous stress) becomes significant while $v$, $\dot{M}$ and $P_{\rm ram}$ (as well as the material stress $\dot{M} R^2 \Omega$) go to zero.
\citet{sun77} used the gas pressure scaling as $P_{\rm gas} \propto R^{-61/20}$ in the ``dead disk'' solution to balance the magnetic pressure to obtain an estimate of the inner radius \citep[see also][]{dav79,dav81}. A description valid for both accretion and propeller cases would employ the equilibrium of  the total pressure $P_{\rm gas}+\rho v^2$ with the magnetic pressure (see e.g.\ \citet{kol+02,rom+02,bes+08}). A conceptually more accurate description of the inner radius of the disks is based on the conservation of angular momentum and  requires the balance of the stresses rather than the pressures \citep{gho79a,gho79b}. The inner radius of the disk will be set at a distance where  magnetic stress  is  balanced by material  and viscous stresses. Equivalently, we can write  it in terms of a torque balance
\begin{equation}
\dot{J}_{\ast} =  - \int_{R_{\rm in}-\Delta R}^{R_{\rm in}} B_{\phi}^{+} B_z R^2 dR
\label{R_general}
\end{equation}
where $B_{\phi}^{+}$ is the toroidal field at the upper plane of the disk, $B_z$ is the poloidal field of the star.
In the \emph{accretion} stage, the first term in Equation (\ref{torque}) --the material torque $\dot{M}_{\rm d}  \Omega R^2$--  dominates and the inner edge of the disk is where the rate at which magnetic stress removes angular momentum is equal to the rate at which material stress brings in \citep{gho79a,gho79b}.
Using $\gamma \equiv B_\phi^+/B_z$, Keplerian rotation $\Omega = \Omega_{\rm K}$ and $B_z = -\mu/R^3$ this gives
$R_{\rm in} = \left(2\sqrt{2}\gamma \right)^{2/7} R_{\rm A}$ \citep{wan96}
which  is similar to the Alfv\'{e}n radius, but conceptually very different. Different physical assumptions about the pitch angle $\gamma$ lead to different formulations of the inner radius of the disk.

In the  \emph{propeller} stage, the second term  in Equation (\ref{torque})--viscous torque $2\pi R^3 \eta  \Omega'$-- dominates and so the inner radius of the disk is the location where
the rate of angular momentum being transferred to the disk is balanced by the rate at which
viscous stresses can transport it outwards
\begin{equation}
\left. 2\pi R^3 \eta \frac{d\Omega}{dR} \right|_{R_{\rm in}}=  - \int_{R_{\rm in}-\Delta R}^{R_{\rm in}} B_{\phi}^{+} B_z R^2 dR, \qquad {\rm if}\,\,  \omega_{\ast}>1 
\label{propeller}
\end{equation}
\citep[see e.g.][]{dan10,dan11,dan12}. 

The torque on left hand side is given by Equation~\eqref{viscous_torque} and the right hand side can be estimated from Equation~\eqref{mag_torque2} as $ \epsilon \mu^2/R_{\rm in}^3$ where $\epsilon = \Delta R /R_{\rm in}\sim 0.01$ and we absorbed $\gamma_{\rm c}$ which is of order unity \citep{gho79a} into the uncertainty in $\epsilon$.  We thus obtain
\begin{eqnarray}
R_{\rm in} &=& \left( \frac{\epsilon \mu^2}{-\dot{J}_{\rm d}} \right)^{1/3} \label{R_in13}  \\
           &=& 10^8\, {\rm cm} \left(\frac{\epsilon}{0.01}\right)^{1/3} \mu_{30}^{2/3} \left(\frac{-\dot{J}_{\rm d}}{10^{34} \, {\rm dyne\, cm}}\right)^{-1/3}.          
\end{eqnarray}
As  $\epsilon = \Delta R/R_{\rm in}$ the above is an implicit solution and the full solution gives
\begin{eqnarray}
R_{\rm in} &=& \left( \frac{\Delta R \mu^2}{-\dot{J}_{\rm d}} \right)^{1/4}
\label{R_in14}  \\
           &=& 10^8\, {\rm cm} \left(\frac{\Delta R}{10^6 \, {\rm cm}}\right)^{1/4} \mu_{30}^{1/2} \left(\frac{-\dot{J}_{\rm d}}{10^{34} \, {\rm dyne\, cm}}\right)^{-1/4}.
\end{eqnarray}

\subsection{Braking Index}

The braking index,
$n \equiv \Omega_{\ast} \ddot{\Omega}_{\ast}/\dot{\Omega}_{\ast}^2$, is an observational handle that could have implications for the pulsar spin-down mechanism.
For a pulsar spinning down purely with the magnetic dipole radiation torque,
\begin{equation}
\dot{J}_{\rm mdr} = -\frac{\mu^2}{c^3} \Omega_{\ast}^3 (1+\sin^2 \zeta)
\label{N_mdr}
\end{equation}
where $\zeta$ is the inclination angle between the rotation and magnetic dipole axis \citep{spi06}
the braking index is 3.  The glitch activity of the Vela pulsar precludes a phase 
coherent measurement of its braking index. A `derived' value of the braking index for the Vela pulsar 
is given as $n=1.4$ \citep{lyn+96} and more recently as $n=1.7$  \citep{esp13}.
We note that these results are very sensitive to the choices in the subtraction of the glitch 
related timing behaviour.
Many authors suggested that the less-than-three braking indices of pulsars can be explained by torques acted by fallback disks in addition to the magnetic dipole radiation torque \citep{mic88,men+01,cal+13}. Action of disk torques may also address the evolution in the $P-\dot{P}$ diagram \citep{alp+01}.

The braking index of a pulsar whose spin down is assisted by the disk torques becomes
\begin{equation}
n =  \frac{3+p \dot{J}_{\rm d}/\dot{J}_{\rm mdr}}{1+ \dot{J}_{\rm d}/\dot{J}_{\rm mdr}}
\end{equation}
where the disk torque $\dot{J}_{\rm d}$ is the sum of the propeller and magnetic torques acted by the disk onto the star and $p \equiv d\ln \dot{J}_{\rm d}/d\ln \Omega_{\ast}$ measures how strongly $\dot{J}_{\rm d}$ depends on $\Omega_{\ast}$.
As the total torque is $\dot{J}_{\rm tot}=\dot{J}_{\rm d}+\dot{J}_{\rm mdr}$ we can eliminate  $\dot{J}_{\rm d}$ to obtain
\begin{equation}
n = p +(3-p)  \frac{\dot{J}_{\rm mdr}}{\dot{J}_{\rm tot}}.
\label{brake_gamma}
\end{equation}
The total torque can be estimated from the period and period derivative as $\dot{J}_{\rm tot}=I \dot{\Omega}_{\ast}=-9.764\times 10^{34}I_{45}$. The magnetic dipole radiation torque is estimated, from Equation (\ref{N_mdr}), as $\dot{J}_{\rm mdr}=-2.43\times 10^{34} \mu_{30}^2 $ where we assumed the inclination angle to be $\xi = 70^\circ$ \citep{rad01} 
and $\mu_{30}$ is the magnetic moment of the neutron star in units of $10^{30}\, {\rm G \, cm^3}$. We thus obtain $\dot{J}_{\rm mdr}/\dot{J}_{\rm tot}=0.25\mu_{30}^2/I_{45}$ and employing the efficient propeller torque \citep{men+01} $p = 1$ Equation (\ref{brake_gamma}) becomes $n=1 + 0.50 \mu_{30}^2/I_{45}$.
 The measured braking index can be explained in this model if $\mu_{30}^2/I_{45} \simeq 1-1.5$. If one employs the torque model in the original propeller model \citep{ill75}  where $p = -1$ Equation (\ref{brake_gamma}) becomes $n=-1 + 1.00 \mu_{30}^2/I_{45}$. In this case $\mu_{30}^2/I_{45} \simeq 2.5-3$ is required to explain the braking index. In the model employed in this work the torque is independent of the spin of the star i.e.\ $p=0$ which gives
\begin{equation}
n =\frac{3}{1+\dot{J}_{\rm d}/\dot{J}_{\rm mdr}} = 3  \frac{\dot{J}_{\rm mdr}}{\dot{J}_{\rm tot}}=0.75\frac{\mu_{30}^2}{I_{45}}
\label{brake_gamma2}
\end{equation}
which then implies that one requires $\mu_{30}^2/I_{45} \simeq 2$ in order to explain the inferred braking index of $\simeq 1.5$. Similarly, using Equations \eqref{N_mdr} and \eqref{mag_torque2} in the equation above
\begin{equation}
n =\frac{3}{1+\frac{\epsilon \gamma_{\rm c}}{1+\sin^2 \xi} \left(\frac{R_{\rm L}}{R_{\rm in}}\right)^3} = \frac{3}{1+0.41 \left(\frac{\epsilon \gamma_{\rm c}}{0.01} \right) \left(\frac{R_{\rm in}}{10^8~{\rm cm}} \right)^{-3} }
\label{brake_gamma3}
\end{equation}
where we assumed the inclination angle to be $\xi = 70^\circ$ \citep{rad01} and employed $R_{\rm L}=4.26\times 10^{8}$~cm. This last equation, using the inner radius and boundary layer width obtained from our spectral fits gives a braking index of $n=2.0 \pm 0.2$ which is somewhat greater than the measured values of $n=1.4$ \citep{lyn+96} and $n=1.7$  \citep{esp13}. Thus we conclude that the disk model can partially explain 
the less-than-3 discrepancy in the case of the Vela pulsar.

\subsection{Outer radius and angular momentum of the disk}
\label{sec-outerradius}

The outer radius of $R_{\rm out} \simeq 1\times 10^{11}$ cm, as obtained from the spectral fit 
in \S~\ref{sec-spectrum}, corresponds to a disk with 
an average specific angular momentum of $\langle j \rangle = 0.45 \sqrt{GM R_{\rm out}} = 1.94\times 10^{18} \, {\rm cm^2 \, s^{-1}}$
(see Appendix~\ref{sec-specific}) where we assumed  $M=1.4\,M_{\odot}$.
This corresponds to a total angular momentum of
\begin{equation}
J_{\rm d} = 3.9\times 10^{46} \, {\rm cm^2 \, s^{-1}} \left( \frac{R_{\rm out}}{10^{11} \, {\rm cm}} \right)^{1/2} \left(\frac{M_{\rm d}}{10^{-5}M_{\odot}} \right)
\end{equation}
which is comparable to the angular momentum of the neutron star $J_{\ast} =7\times 10^{46} I_{45}(P/89 \, {\rm ms})^{-1}$. Formation of fallback disks following the supernova explosion partly addresses where some fraction of the angular momentum of the progenitor goes.

\subsection{Jets and PWN}

The jets in Vela \citep{hel+01,pav+03} and Crab \citep{hes+02} systems were attributed to the presence of fallback disks around these objects \citep{bla04}. A jet indeed is a natural consequence of the presence of a disk given the ubiquity of jets in many types of accreting systems. Due to the long persisting paradigm that Vela and Crab are isolated objects, however, the presence of jets in these systems were more commonly attributed to the PWN \citep{lyu01,tsi02,kom03}.  The strong propeller we infer from the spectral fits to the disk around Vela provides a natural setting for launching jets from this system \citep[see e.g.][]{rom+09,lii+12,ust+06,lov+99} contributing to the mechanisms related to PWN.

\section{Conclusion}
\label{sec-conclude}

Our high resolution X-ray analysis suggest the presence of circumstellar  material, possibly a fallback disk as proposed by \citep{dan+11} who modeled the infrared spectrum by an irradiated passive disk well beyond the light cylinder. As such a disk does not have a stable inner boundary we looked for a disk model protruding the light cylinder. Our modeling of the larger portion of the spectrum including the optical and UV bands with the quiescent disk model of SS77 suggests that Vela and its debris disk form a system in a strong propeller stage. 
The inner radius of the disk is found to be inside the light cylinder and would then be interacting with the magnetosphere. 
The presence of disk torques assisting the magnetic dipole radiation torques partly explains the braking index being substantially less than 3 \citep{men+01}, and may also address the evolution in the $P-\dot{P}$ diagram \citep{alp+01}. If the excess emission in the infrared is associated with the counter-jet of the Vela pulsar as suggested by \citet{zyu+13}, the irradiated fallback disk interpretation of the spectrum in that band would not make sense. Yet the excess in the UV band from the extrapolation of the surface black body is naturally explained in the fallback disk model. Such excess emission is also present in some of the dim thermal neutron stars \citep{kap+09} and may be addressed by the presence of a circumstellar material \citep{ham+09}.

We stress that the  quiescent disk solution of  SS77 and \citet{pri74} in which the mass flux within the disk is halted because of the large propeller torques acted at the inner rim of the disk is not to be confused with the notion of  ``passive disks'' in which the mass flux is reduced due to insufficient ionization such that magnetorotational instability \citep{bal98} can not operate and sustain turbulence. Furthermore, we would like to emphasize that the quiescent disk picture of propelled disks that we present in this work is well different from what had been employed in the fallback disk literature up to now \citep[see e.g.][]{eks03} where the propeller stage was conceived as an episode with substantial mass influx that can not accrete but is totally ejected. The large mass influx leads to large viscous dissipation \citep{per+00} with optically bright disks even in the propeller stage. Our fitting of the spectrum of Vela favours a disk flow with closer resemblance to the  quiescent disk solutions of SS77 and \citet{pri74} where the stopping of the material at the inner boundary halts the mass inflow totally.

Yet we have also seen that the boundary layer luminosity and the viscous luminosity of the Keplerian disk are very small compared to the spin-down luminosity. This indicates that the torque required to explain the viscous dissipation in the disk is much smaller than the torque required to explain the inner radius via Eq.\eqref{R_in13} and the braking index via Eq.~\eqref{brake_gamma2}. To rescue the fallback disk model one might argue that the energy transferred by the magnetosphere of the star is mostly taken by the kinetic energy of the outflow. The presence of an an outflow would imply the presence of same amount of inflowing matter and associated dissipation in the disk which would overwhelm the observed fluxes.  If, however, the disk is advection dominated \citep{nar94,ich77} the luminosity of the disk would be substantially lowered. Note that such advection dominated disk solutions favour the presence of outflows. Advective fallback flow solutions with propeller boundary conditions is beyond the scope of the present work.

\acknowledgements
 KYE acknowledges support from
the scientific and technological council of 
Turkey (TUBITAK) with the project number 112T105.
VVH, RN, and MMH would like to thank the
German national science foundation
(Deutsche Forschungsgemeinschaft, DFG)
for support in the Collaborative Research
Project on Gravitational Wave Astronomy
(SFB TR 7). Based on observations made with the NASA/ESA Hubble Space Telescope, obtained from the Data Archive at the Space Telescope Science Institute, which is operated by the Association of Universities for Research in Astronomy, Inc., under NASA contract NAS 5-26555.
Based on observations made with ESO Telescopes at the La Silla Paranal Observatory under programme ID 66.D-0568.
Based on observations obtained at the Gemini Observatory under program ID GS-2012B-SV-412, which is operated by the
Association of Universities for Research in Astronomy, Inc., under a cooperative agreement
with the NSF on behalf of the Gemini partnership: the National Science Foundation (United
States), the National Research Council (Canada), CONICYT (Chile), the Australian Research
Council (Australia), Minist\'{e}rio da Ci\^{e}ncia, Tecnologia e Inova\c{c}\~{a}o (Brazil) and
Ministerio de Ciencia, Tecnolog\'{i}a e Innovaci\'{o}n Productiva (Argentina).
This research has made use of the SIMBAD database, operated at CDS, Strasbourg, France.
This research has made use of NASA’s Astrophysics Data System Bibliographic Services.
This research has made use of the AstroBetter blog and wiki.
This research has made use of the Scipy and Matplotlib Python libraries.
CG wishes to acknowledge Deutsche Forschungsgemeinschaft (DFG) for
grant MU2695 /13-1. CG also wishes to express special thanks to Donna Keeley for language editing.
We thank M.~Ali Alpar and the anonymous referee for useful suggestions.

\bibliographystyle{apj}

\appendix

\phantomsection \label{Sec:appendix}

\section{Evolution of the fallback disk}

Fallback disks are not tidally torqued as disks in binary systems and can expand freely either to conserve its total angular momentum when matter is being lost from its inner boundary or to take up the extra angular momentum added to the disk by the magnetosphere of the neutron star. In this section we use the self-similar solution given by \citet{pri74} to describe the evolution of the disk in the propeller stage.
 
Evolution of the surface mass density $\Sigma$ in the disk is described by the diffusion equation
\begin{equation}
\frac{\partial \Sigma }{\partial t}=\frac{3}{R}\frac{\partial }{\partial R}%
\left[ R^{1/2}\frac{\partial }{\partial R}\left( \nu \Sigma R^{1/2}\right) %
\right] .
\label{diff1}
\end{equation}
The equation is linear if viscosity, $\nu$, does not depend on $\Sigma$.
In general the viscosity depends on $\Sigma$ and the equation is nonlinear. 

The rest of the disk structure equations are the same as given in \citet{sha73}. Solving these algebraic disk structure equations among themselves one obtains viscosity (and other variables) in terms of $R$ and $\Sigma$ in the form 
\begin{equation}
\nu = C R^p \Sigma^q
\label{viscosity}
\end{equation}
where $C$, $p$ and $q$ are constants determined by the dominant opacity regime and 
pressure  \citep[see e.g.][]{can+90}. This then can be plugged in 
the diffusion equation (\ref{diff1}) to obtain an equation containing $\Sigma$ only. 
For disks in which gas pressure and electron scattering opacity dominates $p=1$ and $q=2/3$, 
and for bound-free opacity $p=15/14$ and $q=3/7$ \citep{can+90}. We assume bound-free opacity 
regime to prevail throughout the disk though different opacity and pressure regimes, depending 
on temperature, are expected to dominate at different locations of the disk. For this opacity regime
\begin{eqnarray}
C &=& \alpha^{8/7}\left(\frac{ 27 \kappa_0}{\sigma_{\rm SB}}\right)^{1/7}\left(\frac{k_B}{\bar{\mu} m_p}\right)^{15/14}(GM)^{-5/14}  \\
  &=& 441.34 \left( \frac{\alpha}{0.1}\right)^{8/7}  \left(\frac{\bar{\mu}}{0.625}\right)^{-15/14} \left(\frac{M}{1.4\,M_{\odot}}\right)^{-5/14}.
\label{CC}
\end{eqnarray}

\subsection{Pringle's Quiescent Disk Solution}

In order to apply self-similarity methods we render the equation dimensionless  via
${\cal R}=R/R_0$, $\tau=t/t_0$ and $\sigma=\Sigma/\Sigma_0$, and define $\nu_0=C R_0^p \Sigma_0^q$. 
Choosing $t_0=4 R_0^2/3\nu_0$ we obtain
\begin{equation}
\frac{\partial \sigma }{\partial \tau }=\frac{4}{{\cal R}}\frac{\partial }{\partial
{\cal R}}\left[ {\cal R}^{1/2}\frac{\partial }{\partial {\cal R}}\left( {\cal R}^{p+1/2} \sigma^{q+1}\right) \right]. 
\label{diff2}
\end{equation}
Self-similarity methods provide two solutions of this equation as first found by 
\citet{pri74}. The first solution corresponds to the accretion regime \citep{can+90,fil+88} and 
is not relevant for our case. The second solution is the quiescent disk solution corresponding 
to the propeller regime \citep{pri91,pri81} and can be written as
\begin{equation}
\sigma({\cal R},\tau) =k_{p,q}{\cal R}_{\rm out}^{-2}
\left( \frac{{\cal R}}{{\cal R}_{\rm out}} \right)
^{-\frac{p+1/2}{q+1}}\left[ 1-\left( \frac{{\cal R}}{{\cal R}_{\rm out}} \right) ^{%
\frac{5}{2}-\frac{p+1/2}{q+1}}\right] ^{1/q}
\label{sol2}
\end{equation}
where 
\begin{equation}
k_{p,q}=\left( \frac{q}{( 4q-2p+4) ( 5q-2p+4)} \right)^{1/q}.
\end{equation}
and
\begin{equation}
{\cal R}_{\rm out}(\tau) = (1+\tau)^{\frac{2}{4q-2p+4}}
\label{R_out}
\end{equation}
is the outer radius of the disk in units of $R_0$, the initial outer radius of the disk.
Note that Equation~(\ref{diff1}) is
symmetric under translations in time and 
we have exploited this to write the original solutions shifted as $\tau \rightarrow \tau + 1$. 

In the non-dimensionalization process we have employed four quantities, $\Sigma_0$, $R_0$, $t_0$ and $\nu_0$, related by two equations $\nu_0=CR_0^p\Sigma_0^q$ and $t_0=4R_0^2/3\nu_0$. This means we are free to attribute any value for two of the variables e.g.\ $\Sigma_0$ and $R_0$. We fix these two quantities in terms of the initial mass and angular momentum of the disk as shown in \citet{ert+09} for the accretion solution.

\subsection{Mass of the disk}

The mass of the disk is
\begin{equation}
M_{\mathrm{d}}=\int_0^{R_{\rm out}} \Sigma \cdot 2\pi RdR = 2\pi R_0^2 \Sigma_0 \int_0^{{\cal R}_{\rm out}} \sigma   {\cal R}d{\cal R}.
\label{M_d}
\end{equation}
Here $R_{\rm out}$ is the freely expanding outer boundary of the disk.
The lower limit of the integrals is zero as the solutions extend to the origin. 
The mass and angular momentum of the disk is mostly carried by the outer parts and, as long as this outer radius is much greater than the inner radius, the integrals provide the the mass and angular momentum of the disk accurately.
Using the solution given in Equation (\ref{sol2}) in Equation (\ref{M_d}) and defining $x={\cal R}/{\cal R}_{\rm out}$ we find
\begin{equation}
M_{\mathrm{d}}= 2\pi R_0^2 \Sigma_0 \gamma_m
\label{M_d2}
\end{equation}
where
\begin{eqnarray}
\gamma_m &=& k_{p,q} \int_0^{1} 
x^{1-\frac{p+1/2}{q+1}}
\left( 1-x  ^{\frac{5}{2}-\frac{p+1/2}{q+1}}\right)^{1/q}  dx \\
&=& k_{p,q} \frac{2\left( q+1\right) }{5q+4-2p}{\rm Beta}\left( \frac{q+1}{q},\frac{4q-2p+3}{5q+4-2p}\right)
\label{gamma_m}
\end{eqnarray}
Numerical value of $\gamma_m$ for bound-free opacity and electron-scattering opacity dominated disks are given in Table~\ref{tab:param}. Note that in this solution $\dot{M}_{\rm d}=0$ hence the solution is the time-dependent version of the quiescent disk solution given by \citet{sun77} in the steady-state.
Note that $\dot{M}_{\rm d}=0$ is a rather extreme boundary condition and recently, \citet{eks12} presented an approximate solution that can have finite mass flux in the propeller stage. 
Here we continue with the $\dot{M}_{\rm d}=0$ solution which allows an exact treatment in the following. 

\subsection{Angular momentum of the disk}
\label{sec-specific}

The angular momentum of the disk is
\begin{equation}
J_{\mathrm{d}}=\int_0^{R_{\rm out}} R^2 \Omega_{\mathrm K} \Sigma \cdot 2\pi RdR = 
2\pi \Sigma_0  \sqrt{GMR_0} R_0^2 \int_0^{{\cal R}_{\rm out}}   \sigma   {\cal R}^{3/2}d{\cal R}
\label{J_d}
\end{equation}
where ${\cal R}_{\rm out}$ is given in Equation (\ref{R_out}).
Using the solution given in Eqn. (\ref{sol2}) in this equation and defining $x={\cal R}/{\cal R}_{\rm out}$
we get
\begin{equation}
J_{\mathrm{d}} = 
2\pi R_0^2 \Sigma_0  \sqrt{GMR_0} {\cal R}_{\rm out}^{1/2}\gamma_{\ell} 
\label{J_d2}
\end{equation}
where 
\begin{equation}
\gamma_{\ell} = k_{p,q} \int_0^{1}   x^{\frac32 - \frac{p+1/2}{q+1}} 
 \left( 1-x^{\frac{5}{2}-\frac{p+1/2}{q+1}}\right) ^{1/q}  dx= k_{p,q} \frac{2q}{5q-2p+4}
\end{equation}
(see Table~\ref{tab:param} for numerical values). 
The angular momentum can be written as
\begin{equation}
J_{\mathrm{d}} = 
2\pi R_0^2 \Sigma_0  \sqrt{GMR_{\rm out}} \gamma_{\ell} = M_{\rm d}  \sqrt{GMR_{\rm out}} \frac{\gamma_{\ell}}{\gamma_m}
\label{J_d3}
\end{equation}
where 
\begin{equation}
\frac{\gamma_m}{\gamma_{\ell}} =\frac{q+1}{q}   {\rm Beta}\left( \frac{q+1}{q},\frac{4q-2p+3}{5q-2p+4}\right)
\end{equation}
and $\gamma_m/\gamma_{\ell}=1.3326/0.59925=2.224$ for bound-free opacity.
The average specific angular momentum is then
\begin{equation}
\langle j \rangle = 0.45 \sqrt{GM R_{\rm out}} = 1.94\times 10^{18} \, {\rm cm^2 \, s^{-1}} \left(  \frac{M}{1.4\, M_{\odot}} \right)^{1/2} \left( \frac{R_{\rm out}}{10^{11} \, {\rm cm}} \right)^{1/2}.
\end{equation}
for a disk where bound-free opacity dominates.

\subsection{Evolution of the disk in terms of initial mass and angular momentum}
\label{m_0_j_0}

At $t=0$ we have from Eqn. (\ref{R_out}) that ${\cal R}_{\rm out}=1$ and so from Eqn. (\ref{J_d2}) the initial angular momentum of the disk is
\begin{equation}
J_0 = 2\pi R_0^2 \Sigma_0  \sqrt{GM R_0} \gamma_{\ell}.
\label{J_0}
\end{equation}
As the mass of this disk in this propeller solution is constant ($M_0=M_{\rm d}$) we have from Eqn.(\ref{M_d2}) that
\begin{equation}
M_0 = 2\pi R_0^2 \Sigma_0 \gamma_m.
\label{M_0}
\end{equation}
From the latter two equations
\begin{equation}
R_0 = \frac{(J_0/M_0)^2}{GM}  \left(\frac{\gamma_m}{\gamma_{\ell}} \right)^2 
=2.66 \times 10^{10}\, {\rm cm} \left( \frac{\langle j_0 \rangle}{10^{18} \, {\rm cm^2\, s^{-1}}}\right)^2  \left(\frac{M}{1.4\, M_{\odot}}\right)^{-1}
\label{R_0}
\end{equation}
where $\langle j_0 \rangle=J_0/M_0$ is the average specific angular momentum of the disk.
Once $R_0$ is determined, one can find $\Sigma_0$ from  Eqn. (\ref{M_0}) as
\begin{equation}
\Sigma_0 = \frac{M_0}{ 2\pi R_0^2 \gamma_m} =  1.67\times 10^{10} {\rm g\, cm^{-2}} \left(\frac{M_0}{10^{-5}\, M_{\odot}}\right)  \left( \frac{\langle j_0 \rangle}{10^{18} \, {\rm cm^2\, s^{-1}}}\right)^{-4}  \left(\frac{M}{1.4\, M_{\odot}}\right)^{2}
\label{Sigma_0}
\end{equation}
and $\nu_0= C R_0^p \Sigma_0^q$ as
\begin{equation}
\nu_0 = 1.57 \times 10^{18} {\rm cm^2\, s^{-1}} \left( \frac{\alpha}{0.1}\right)^{8/7}  \left(\frac{\bar{\mu}}{0.625}\right)^{-15/14} \left(\frac{M}{1.4\,M_{\odot}}\right)^{-3/14} \left(\frac{M_0}{10^{-5}\, M_{\odot}}\right)^{3/7} \left( \frac{\langle j_0 \rangle}{10^{18} \, {\rm cm^2\, s^{-1}}}\right)^{-12/7}.
\label{nu_0}
\end{equation}
From these one can find $t_0 = 4R_0^2/3\nu_0$ as
\begin{equation}
t_0 = 6.0\times 10^2\, {\rm s} 
 \left( \frac{\alpha}{0.1}\right)^{-8/7}  \left(\frac{\bar{\mu}}{0.625}\right)^{15/14}
  \left(\frac{M}{1.4\,M_{\odot}}\right)^{-25/14}
  \left(\frac{M_0}{10^{-5}\, M_{\odot}}\right)^{-3/7} 
\left(\frac{\langle j_0 \rangle}{10^{18} \, {\rm cm^2\, s^{-1}}}\right)^{40/7}.
\label{t_0}
\end{equation}
Note that $t_0$ depends very strongly on $\langle j_0 \rangle$; e.g.\ for  $\langle j_0 \rangle=10^{19}\, {\rm cm^2\, s^{-1}}$ and  $\langle j_0 \rangle=10^{20}\, {\rm cm^2\, s^{-1}}$ one finds $t_0=9.9$ years and $t_0 = 5.1\times 10^6$ years, respectively.
Another important property is the inverse dependence of $t_0$ on the initial mass of the disk $M_0$; a  disk with a larger mass 
with the same average specific angular momentum will evolve more rapidly because, in order to have the same specific angular momentum, more mass of the disk should be placed at smaller orbits.

The Eqns. \eqref{J_d2} and \eqref{J_0} together with \eqref{R_out} imply that
\begin{equation}
J_{\rm d} = J_0 \left(1 + \frac{t}{t_0} \right)^{1/(4q-2p+4)}
\end{equation}
where, for bound-free opactiy ($p=15/14$ and $q=3/7$), the exponent is 7/25. Thus the angular momentum of the disk increases by the torque acting from the origin. From the above one finds the torque as
\begin{equation}
\dot{J}_{\rm d} = \dot{J}_0 \left(1 + \frac{t}{t_0} \right)^{-\beta}, \qquad \beta = 1- \frac{1}{4q-2p+4}.
\label{Jdot_d_analytical}
\end{equation}  
Here $\beta =18/25$ for bound-free opacity and 
\begin{equation}
\dot{J}_0 = \frac{J_0}{(4q-2p+4)t_0} = \frac{7J_0}{25t_0}
\end{equation}
Thus, the torque decreases in time as the disk diffuses.

\subsection{Energy of the disk and its rate of change}

 The energy density of a ring of mass $dm = 2\pi \Sigma RdR$ would be%
\begin{equation}
d e =\frac{1}{2}\left( \Omega_{\rm K}R\right)^2 dm-\frac{G M dm}{R} = -\frac{G M dm}{2R} = -\pi G M \Sigma  dR
\end{equation}%
where we ignored the kinetic energy associated with the radial motion as it is much smaller than the one associated with the toroidal motion. The total mechanical energy of the disk  $E= \int d e = -\pi G M \int \Sigma  dR$ is then
\begin{equation}
E=  -\pi GM \Sigma_0 R_0 \int_{\mathcal{R}_{\rm in}}^{\mathcal{R}_{\rm out}}\sigma d\mathcal{R}
\end{equation}
where $\mathcal{R}_{\rm in} = R_{\rm in}/R_0$. In this case we employed a finite inner bound to the disk as the contribution to the total energy could be important and leads to the divergence of the integral for the bound-free opacity regime. Using Eqn.\eqref{sol2} and referring $x={\cal R}/{\cal R}_{\rm out}$
\begin{equation}
E =  -\frac{GM}{2 \mathcal{R}_{\rm out}}2\pi \Sigma_0 R_0 k_{p,q} I_{p,q} \qquad I_{p,q}=\int_\delta^1 x^{-\frac{p+1/2}{q+1}}
    \left( 1- x ^{\frac{5}{2}-\frac{p+1/2}{q+1}}\right)^{1/q} dx
\end{equation}
where $\delta =R_{\rm in}/R_{\rm out}$. For $\delta \ll 1$ the integral gives $I_{\rm bf}=10\delta^{-1/10}-11.264$ for the bound-free opacity regime and $I_{\rm es}=-10\epsilon^{1/10}+9.25$ for the electron scattering opacity regime. From our spectral fits $\delta \simeq 0.001$ and the integrals have the values $I_{\rm bf} = 8.68$ and $I_{\rm es}=4.24$, respectively. From the above it is possible to calculate the rate of increase of the mechanical energy of the disk by the torque acted from the origin:
\begin{equation}
\dot{E} = \gamma_{e} \frac{GMM_0}{R_0 t_0} \left(1 + \frac{t}{t_0}\right)^{-1-2/(4q-2p+4)}
\end{equation} 
where $\gamma_{e}= k_{p,q}I_{p,q}/\gamma_m (4q-2p+4)$. For bound-free opacity we obtain $\gamma_{e}=5.1$ and so
\begin{equation}
\dot{E}_{\rm bf} = 5.1 \frac{GMM_0}{R_0 t_0} \left(1 + \frac{t}{t_0}\right)^{-39/25} = 5.1 \frac{GMM_0}{R_0 t_0} \left( \frac{\dot{J}_{\rm d}}{\dot{J}_0}\right)^{39/18}
\end{equation} 
where we referred Eqn.~\eqref{Jdot_d_analytical} in the last step.

\newpage

\begin{table}
  \caption{Observation epochs}
  \begin{tabular}{@{}lllccc@{}}
  \hline
      
 Date					& Telescope 			& Instrument 	 		& Filter 		& Exposure Time [s] 						& Zero Point [mag] 	\\
 \hline
 
	1997-06-30	& HST							& WFPC2						& 555\,W		&	2 $\times$ 1300								&	24.68\,$\pm$\,0.02				\\
	1998-01-02	& HST							& WFPC2						& 555\,W		&	2 $\times$ 1000								&	24.68\,$\pm$\,0.02				\\
	1999-06-30	& HST							& WFPC2						& 555\,W		&	2 $\times$ 1300								&	24.68\,$\pm$\,0.02				\\
	2000-01-15	& HST							& WFPC2						& 555\,W		&	2 $\times$ 1300								&	24.68\,$\pm$\,0.02				\\
	2000-07-05	& HST							& WFPC2						& 555\,W		&	2 $\times$ 1300								&	24.68\,$\pm$\,0.02				\\						
	2000-03-15	& HST							& WFPC2						& 675\,W		&	2 $\times$ 1300								&	24.88\,$\pm$\,0.04				\\
	2000-03-19	& HST							& WFPC2						& 814\,W		&	2 $\times$ 1300								&	25.03\,$\pm$\,0.02				\\
	2000-12-14	& VLT (UT3)				&	ISAAC						& J$_{s}$		&	39 $\times$ 45								& 24.81\,$\pm$\,0.04				\\
	2001-01-05	& VLT (UT3)				&	ISAAC						& H					&	67 $\times$ 13								& 24.56\,$\pm$\,0.07				\\
	2013-01-30 	& Gemini-South		& GSAOI + GeMS 		& K$_{s}$		& 19 $\times$ 100  							& 25.3318\,$\pm$\,0.0063		\\
 
\hline\end{tabular}
\label{tab: obs}
\end{table}

\pagebreak
\newpage

\begin{table}
  \caption{Photometric parameters and results of our NIR and VIS photometry. The apparent magnitudes are de-reddened and corrected for atmospheric extinction (when applicable). The apparent magnitudes of the HST measurements are in the STmag magnitude system, while the ground based NIR magnitudes are in the standard Vega based system.}
  \begin{tabular}{@{}cccccc@{}}
  \hline
      
 Date				& Filter 	& Appar. Magnitude [mag] 		& $\nu_{\rm eff}$ [Hz] 					& Width [Hz] 			& $\nu F_\nu$ [erg s$^{-1}$ cm$^{-2}$]	\\
 \hline
 1997-06-30 & 555\,W	& 23.376\,$\pm$\,0.059			&	$5.681 \times	10^{14}$		&	$1.714	\times	10^{14}$	& $8.01^{+0.47}_{-0.44} \times	10^{-15}$\\
 1998-01-02 & 555\,W	& 23.455\,$\pm$\,0.067			&														&													& $7.45^{+0.49}_{-0.47} \times	10^{-15}$\\
 1999-06-30 & 555\,W	& 23.315\,$\pm$\,0.061			&														&													& $8.48^{+0.51}_{-0.48} \times	10^{-15}$\\
 2000-01-15 & 555\,W	& 23.434\,$\pm$\,0.062			&														&													& $7.60^{+0.46}_{-0.44} \times	10^{-15}$\\
 2000-07-05	& 555\,W	& 23.304\,$\pm$\,0.058			&														&													& $8.56^{+0.49}_{-0.47} \times	10^{-15}$\\
 averaged		& 555\,W	& 23.38\,$\pm$\,0.10				&	$5.681 \times	10^{14}$		&	$1.714 \times	10^{14}$	& $7.98^{+0.80}_{-0.73} \times	10^{-15}$\\
 2000-03-15	& 675\,W	& 23.995\,$\pm$\,0.084			&	$4.445 \times	10^{14}$		&	$0.804 \times	10^{14}$	& $5.99^{+0.49}_{-0.45} \times	10^{-15}$\\
 2000-03-19	& 814\,W	& 24.164\,$\pm$\,0.052			&	$3.611 \times	10^{14}$		&	$1.097	\times	10^{14}$	& $6.12^{+0.31}_{-0.30} \times	10^{-15}$\\
 2000-12-14	& J$_{s}$	& 22.31\,$\pm$\,0.11				&	$2.415 \times	10^{14}$		&	$0.299	\times	10^{14}$	& $4.21^{+0.45}_{-0.41} \times	10^{-15}$\\
 2001-01-05	& H				& 21.76\,$\pm$\,0.17				&	$1.837 \times	10^{14}$		&	$0.319	\times	10^{14}$	& $3.14^{+0.54}_{-0.46} \times	10^{-15}$\\
 2013-01-30 & K$_{s}$	& 21.748\,$\pm$\,0.067			&	$1.408 \times	10^{14}$		& $0.202 \times		10^{14}$	& $1.89^{+0.12}_{-0.11} \times	10^{-15}$\\
 
\hline\end{tabular}
\label{tab: phot}
\end{table}

\pagebreak
\newpage

\begin{table}
\centering
\caption[]{Line parameters of the narrow absorption feature at 0.57~keV derived from the co-added {\it RGS} data. We list the central energy ($\mathrm{E_{line}}$), the optical depth ($\tau$) and the equivalent width (EW) calculated for different models of the ISM ({\it phabs} versus {\it TBabs}, \citet{wil+00}) and for both methods of data analysis (co-adding all spectra versus taking the intensity profile of the source into account). Note that the significance of the absorption feature (given in parenthesis) depends on the data analysis methods. All errors denote the 90\% confidence interval.}
\label{fitvalRGS}
\begin{tabular}{ccc}
\hline
ISM model & all co-added & co-added with intensity profile\\
\hline
{\it phabs} & $\mathrm{E_{line}=568.9_{-3.7}^{+3.3}~eV}$ & $\mathrm{E_{line}=566.5_{-4.2}^{+4.1}~eV}$ \\
            & $\mathrm{\tau=1.69_{-0.64}^{+0.69}\times10^{-3}}$     & $\mathrm{\tau=2.90_{-0.81}^{+0.57}\times10^{-3}}$ \\
            & $\mathrm{EW=-2.0\pm1.2}$  (equals 2.8$\sigma$)  & $\mathrm{EW=-4.2\pm1.3}$ (equals 5.2$\sigma$)\\
\hline
{\it TBabs} & $\mathrm{E_{line}=568.1_{-3.9}^{+3.8}~eV}$ & $\mathrm{E_{line}=565.8_{-4.4}^{+4.0}~eV}$ \\
           & $\mathrm{\tau=1.71_{-0.66}^{+0.57}\times10^{-3}}$     & $\mathrm{\tau=2.83_{-0.74}^{+0.55}\times10^{-3}}$ \\
           & $\mathrm{EW=-1.77\pm0.67}$  (equals 4.2$\sigma$)  & $\mathrm{EW=-3.8\pm1.4}$ (equals 4.2$\sigma$)\\
\hline
\end{tabular}
\end{table}

\begin{table}
\caption{Numerical parameters for disk evolution}
\begin{tabular}{lccccccc}
\hline
Opacity regime & $p$ & $q$ & $k_{p,q}$ & $\gamma_m$ & $\gamma_{\ell}$ & $\beta$  & $\gamma_e$  \\
\hline
Electron scattering & 1 & 2/3 & $4.3838\times 10^{-3}$ & $2.0057\times 10^{-3}$ & $1.0960\times 10^{-3}$ & 11/14 & 1.986  \\
Bound-free & 15/14 & 3/7 & $2.7965\times 10^{-4}$ & $1.3326\times 10^{-4}$ & $0.59925\times 10^{-4}$ & 18/25 & 5.100 \\
\hline
\end{tabular}
\label{tab:param}
\end{table}

\end{document}